\documentclass[aps,twocolumn]{revtex4}

\usepackage[dvips]{graphicx}

\newcommand{\ket}[1]{\left| {#1} \right>}

\newcommand{\beq}{\begin{equation}}
\newcommand{\eeq}{\end{equation}}

\begin{document}

\title{Fault-Tolerant Logical Gate Networks for CSS Codes}
\author{Andrew M. Steane and Ben Ibinson\\
\small Centre for Quantum Computation,
Department of Atomic and Laser Physics, University of Oxford, \\
\small Clarendon Laboratory, Parks Road, Oxford OX1 3PU, England.
}

\date{\today}

\begin{abstract}
Fault-tolerant logical operations for qubits encoded by CSS codes
are discussed, with emphasis on methods that  apply to codes of high rate, encoding
$k$ qubits per block with $k>1$. It is shown that  the logical qubits within a given
block can be prepared by a single recovery operation in any state whose
stabilizer generator separates into $X$ and $Z$ parts. Optimized methods
to move logical qubits around and to achieve controlled-not and Toffoli gates
are discussed. It is found that  the number of time-steps required to
complete a fault-tolerant quantum computation is the same when $k>1$ as when $k=1$.
\end{abstract}


\maketitle

\section{Introduction}

Fault tolerant quantum computation is quantum computation of high
fidelity carried out with physical qubits and operations that  are
noisy and imperfect. `Fault tolerance' covers a variety of
concepts, but there are three main ones: (generalized) geometric
or adiabatic phases, composite pulses, and quantum error correction
(QEC). This paper is concerned purely with the latter.

The main ideas for fault-tolerant universal quantum computation on
encoded states were introduced by Shor \cite{96:Shor}. Two aspects
have to be considered: the error correction or recovery process,
which uses a noisy quantum network, and the implementation of
quantum gates to evolve the logical state of the machine. This
paper is concerned purely with the latter task, but we will study
methods in which the two aspects are to some extent merged.

The present work builds on
a series of ideas that  were established as follows. Shor's
seminal work \cite{96:Shor} discussed CSS codes encoding a single qubit per
block. It established such central concepts as the use of
ancilliary entangled states that  are partially verified,
repetition of syndrome measurements, and a discrete universal set
of logical operations. DiVincenzo and Shor \cite{96:DiVincenzo}
generalised the fault-tolerant syndrome measurement protocol to
any stabilizer code, and Steane \cite{97:SteaneA} discovered the more
efficient technique of using prepared logical zero states to
extract syndromes, which will be adopted in this paper.

Gottesman \cite{98:GottesmanA} discovered fault-tolerant universal
methods that  can be applied to all stabilizer codes. The main new
ingredient is to use measurements of observables in the Pauli
group, combined with preparation of `cat' states, to achieve
desired operations. Teleportation in particular is used to extract
an individual logical bit from one block and place it in another.
Steane \cite{99:SteaneB} showed that  the measurements of Pauli
observables required in Gottesman's methods can be absorbed into
the syndrome measurement, so that  they are achieved at close to
zero cost.

The important concept of `teleporting a gate' or teleporting
qubits `through' a gate was introduced by Nielsen and Chuang
\cite{97:Nielsen} and applied to fault-tolerant gate constructions
by Gottesman and Chuang \cite{99:GottesmanB}.

In this paper we study methods for quantum
codes encoding more than one qubit per block.
We introduce extensions and generalisations
of the ideas just listed, and identify networks requiring the
least computation resources to perform a given operation.
One interesting result is that  the number of time steps
required to complete a logical algorithm is the same for
$k=1$ and $k > 1$, where $k$ is the number of logical qubits per block.
This is because the methods allow much of the required processing to
take place ``off-line'', without interrupting the evolution of the computer.
The ``off-line'' operations involve the preparation of ancilliary qubits in
specific states, and the transfer of logical qubits to otherwise empty
blocks by teleportation.

The paper is organised as follows.
Section \ref{s:term} introduces terminology and notation. Section \ref{s:U}
lists some ways to achieve a universal set of fault-tolerant operations.
Section  \ref{s:measure} then presents our first main result (theorem 1
and its corollary). This is
an extension of a theorem in \cite{99:SteaneB}, it shows that  CSS-encoded
qubits can be fault-tolerantly prepared in a useful class of states by
use of a single recovery operation. We also discuss how to simplify
some more general state-preparations by decomposing
stabilizer operators into simpler components.

Section \ref{s:toolbox} gives a set of basic operations for CSS
codes. The main aim is to discuss the transfer
and teleportation operations whose use for manipulating bits encoded by
stabilizer codes was proposed by Gottesman \cite{98:GottesmanA}.
We list the constructions and present the most efficient
implementation of teleportation between blocks.
We use theorem 1 to avoid the need to prepare `cat' states
for preparing and measuring states, including states in the Bell basis
of encoded qubits.

Sections \ref{s:cnot} and \ref{s:Toff} discuss implementation of the
controlled-not and Toffoli gates respectively, between qubits encoded
in the same block.

\section{Terminology and notation}  \label{s:term}

The following notation will be adopted. The single-qubit operators
$X$, $Y$ and $Z$ are the Pauli operators $\sigma_x$, $\sigma_y$
and $\sigma_z$, respectively, (it will be convenient to define $Y$
so that  it is Hermitian, not real as is sometimes chosen in QEC
discussions). We use $H$ for the single-qubit Hadamard operation
and $S$ for the rotation about the $z$ axis through $\pi/2$ (phase
shift of $\left| {1} \right>$ by $i$). Thus $S^2=Z$ and $(HSH)^2 =
X$. The general phase shift of $\left| {1} \right>$ by
$\exp(i\phi)$ will be written $P(\phi) $, so $S=P(\pi/2)$,
$Z=P(\pi)$, etc.

A controlled $U$ operation is written $^C\! U$, so for example
$^C\! X$ is controlled-not, and $T \equiv \, ^{CC}\!\! X$ is the
Toffoli gate.

The logic gate hierarchy introduced in \cite{99:GottesmanB} is defined
recursively by
\beq
{\cal C}_j \equiv \{ U \, | \, U {\cal C}_1 U^{\dagger} \subset
{\cal C}_{j-1} \},            \label{Cs}
\eeq
where ${\cal C}_1$ is the Pauli group (the set of tensor products of Pauli
operators, including the identity $I$ and $iI$). Each ${\cal C}_j$ contains
${\cal C}_{j-1}$. $P(\pi / 2^j) \in {\cal C}_{j+1} \setminus {\cal C}_j$
where $\setminus$ denotes the set difference.
The Clifford group is ${\cal C}_2$ in
the heirarchy (\ref{Cs}). By the definition of ${\cal C}_2$, this
group is the normalizer of the Pauli group. It is generated by
$\{H, S, ^C\!\! X\}$ \cite{98:Knill,99:GottesmanB}. 

All operators are understood to act on the logical, i.e. encoded
qubits (operations on the physical qubits are discussed in the appendix).
A {\em blockwise} operation is defined to be one such that  the
relevant operator acts on each of the logical qubits in a given block, or
each corresponding pair in two blocks in the case of 2-qubit
operators (blockwise action of 3- or more-bit operators will not
arise in the discussion).

We define an operation to be `fault tolerant' if it does not cause errors in
one physical qubit to propagate to two or more qubits in any one
block. The fault tolerance of the operations used in the networks
to be discussed is proved in the appendix.

A block of $n$ physical qubits stores $k$ logical qubits. The
notation ${M}_{u}$, where $u$ is an $k$-bit binary word, means a
tensor product of single-qubit $M$ operators acting on those
logical qubits identified by the 1s in $u$ (for example
${X}_{101}={X}\otimes {I}\otimes {X}$). The letters $u,v,w,x,y,z$
when used as a subscript or inside a ket symbol (as in
$\ket{x}_L$) always refer to binary words. When we wish to treat a
list of operators such as $\{ M_i,\;i=1 \dots k \}$ then the
letters $i,j,r,p$ are used as subscripts.

The notation ${X}^i \equiv {X}_{2^{k-i}}$ or ${Z}^i \equiv {Z}_{2^{k-i}}$,
where $i$ is a number running from 1 to $k$, means a single operator applied to the $i$'th
logical bit in a block. For example $X^2 \equiv X_{01000}$ for $k=5$; N.B. no powers
(greater than 1) of Pauli operators appear anywhere in this paper.

\subsection{Computational resources} \label{s:resource}

Most of the computational resources of the physical computer are
dedicated to the QEC networks. The complete network to recover
($\equiv$ error-correct) a single block involves $\sim n d^2$ physical gates
\cite{03:Steane}, where $d$ is the minimum distance of the code,
whereas the operations acting in between recoveries of a given
block typically only involve $n$ physical operators (one for each
physical bit in the block).
To assess the resources of the networks to
be described we will therefore primarily count blocks and recoveries.

Whenever a single block is recovered, all are, because the
duration of the recovery network is assumed to be long enough that 
even `resting' blocks accumulate significant memory errors. We
allow at most one set of gates connecting different blocks between
successive recoveries, to prevent avalanches of errors. However,
we allow combinations of twin- and single-block
operations, such as $^C\!\! X$ followed by $H$, without requiring a
further recovery. We define one `time step'
to be the interval between the completion of one recovery, and the
completion of the next. The `area' of a network is defined to be
the product (number of blocks) $\times$ (number of time steps).

Measurement of logical bits, and preparation of logical bits in
required states, is absorbed as much as possible into the recovery
operations as described in section \ref{s:measure}.

Most of the operations on the computer are either
measurements absorbed into recoveries or
a physical gate applied once to each bit in a block
or pair of blocks (so-called `transversal' application of a gate).
We will
treat in this paper the case where the QEC encoding
is a CSS code based on a doubly even classical code, such that 
fault-tolerant Clifford group gates are relatively
straightforward (see section \ref{s:toolbox}) but the members of ${\cal C}_3$
(including the Toffoli gate and $^C\!\! S,\;P(\pi/4)$) are not. To implement
the latter, we adopt
Shor's method of preparing a block of $n$ physical bits in
the `cat' state $\ket{0^{\otimes n}} + \ket{1^{\otimes n}}$ and
using it to measure Clifford group observables such as blockwise
$^C\!\! X$ on encoded bits. This method is fault-tolerant, but it is
an undesirable element because the noise associated with
preparing the cat states and connecting them to the data qubits
is larger than that  of a single transversal gate. Therefore
we will aim to keep the use of such cat states to a minimum.

We distinguish between `offline' and `online' parts of the networks
to be discussed. The `online' parts
are so called because they involve operations on the logical data qubits of the
computer, and therefore can only take place at the correct moment in the
algorithm being computed. The `offline' parts are state preparations 
which can take place at any time prior to when they are needed, and
operations to move passive qubits (i.e. those not immediately involved
in a logical gate) around in order to conserve memory blocks.
The offline parts can proceed in
parallel with other operations of the computer as long as there are
sufficient
spare blocks available, but the computer's algorithm cannot be
evolved further while the online part of a given step is
completed, because the algorithm (in all but rare instances)
requires the logical operations to take place sequentially.
This means that  when considering the computation resources required
for a given network, {\em the most important cost measure is the duration
of the online part}.

In the methods to be discussed, it often happens that  data qubits are
moved from one block to
another in order to make it possible to apply logical operations
to them. At any given moment, most blocks in the computer act as memory, and a few act
as an `accumulator' where the logical operations take place.
The movement of memory qubits too and from the accumulator is
intermediate between `offline' and `online'. For, suppose a data bit has been
moved to an accumulator block and a logical operation
has just been applied to it. In order to free the
accumulator for further use, the bit must be moved out again.
If this bit were required in the next logical operation, however, then it
is usually possible to apply the logical operation straight away, and
move it afterwards. If the bit were not required,
then the operation to move it back into memory could proceed
offline, as long as there is another accumulator block available
to allow the next logical gate to proceed at the same time.
Therefore we will count each operation to move qubits from memory to
accumulator as online, and operations to move them back to memory
as offline.

\section{Universal sets}  \label{s:U}

In this section we will consider universal sets of quantum gates
for which fault-tolerant constructions have been put forward.

For operations on bare qubits, the most commonly considered
universal set of quantum gates is $\{ U(\theta, \phi), ^C\!\! X \}$
where $U(\theta, \phi)$ is a rotation of a single qubit through $\theta$
about an axis in the $x-y$ plane specified by $\phi$.
However, this is not a useful set to consider for the purpose of
finding fault-tolerant gates on encoded qubits, because $U(\theta,
\phi)$ is not readily amenable to fault-tolerant methods.

Several different proposals for fault-tolerant universal sets have
been put forward. All involve the Clifford group. The Clifford group is not
sufficient for universal quantum computation, nor even for useful
quantum computation, since it can be shown that  a quantum computer
using only operations from the Clifford group can be efficiently simulated
on a classical computer \cite{98:GottesmanB,Bk:Nielsen}. To complete the set a
further operator must be added, and it can be shown \cite{96:Shor,Bk:Nielsen}
that  an operator in ${\cal C}_3 \setminus {\cal C}_2$ suffices.

\begin{enumerate}
\item  Shor \cite{96:Shor} proposed adding the Toffoli gate, making the universal
set $\{H,S,^C\!\! X,T\}$ (or $\{R,S,^C\!\! X,T\}$ which is equivalent
since $R=HS^{2}$). Obviously, $^C\!\! X$ can be obtained from $T$, but this
does not reduce the set since Shor's method to obtain $T$ assumes
that  $^C\!\! X$ is already available.

\item $\{H,S,^C\!\! X,^C\!\! S\}$ was considered for example by
Knill, Laflamme and Zurek \cite{96:KnillB}. This is similar to (1)
because $^C\!\! S$ and $^C\!\! X$ suffice to produce $^{CC}\!\!Z$,
which with $H$ makes $^{CC}\!\!X=T$.

\item The same authors \cite{96:KnillB} also considered $\{S,^C\!\! X,^C\!\! S\}$ together
with the ability to prepare the encoded (or `logical') states
$\left| {+}\right\rangle _{L}\equiv \left( \left| {0}\right\rangle
_{L}+\left| {1}\right\rangle _{L}\right) /\sqrt{2}$,
$\left| {-}\right\rangle _{L}\equiv \left( \left| {0}\right\rangle _{L}-\left| {1}%
\right\rangle _{L}\right) /\sqrt{2}$. This can be shown to be
sufficient since preparation of $\left| {\pm }\right\rangle _{L}$
together with $S$ and $X$ can produce $H$, and the rest follows as
in (2).

\item $\{ H, S, \,^C\!\! X, P(\pi/4) \}$ is the `standard set'
discussed by Nielsen and Chuang \cite{Bk:Nielsen}.

\item  Knill {\em et al.} \cite{98:Knill} proposed $\{H,S,^C\!\! X\}$
combined with preparation of $\left| {\pi /8}\right\rangle
_{L}=\cos (\pi /8)\left| {0}\right\rangle _{L}+\sin (\pi /8)\left|
{1}\right\rangle _{L}$. The latter is prepared by making use of
the fact that  it is an eigenstate of $H$, and once prepared is
used to obtain a $^C\!\! H$ operation, from which the Toffoli gate can
be obtained.

\item  Gottesman \cite{98:GottesmanA} showed that  $^C\!\! X$, combined with the
ability to measure $X,Y$ and $Z$, is sufficient to produce any
operation in ${\cal C}_2$. The universal set
is completed by an operation in ${\cal C}_3 \setminus {\cal C}_2$ such as $T$.

\item Shi \cite{02:Shi} proved that  $\{ H,T \}$ is universal; some
further insights are given by Aharonov \cite{03:Aharonov}.
\end{enumerate}

Many of these methods are summarized and explained in
\cite{Bk:Nielsen}, where the proof of universality and the
efficiency of approximating a continuous set with a discrete one
(Solovay-Kitaev theorem) is also discussed.

(1) is a useful starting point and we will use it in
this paper, but generalized to $[[n,k,d]]$
codes storing more than one qubit per block. Similar methods
apply to (2) and (4).
A generalization of the ideas of Knill {\em et al.} used for (2)
is given in the appendix; however, the codes for which it works
turn out to be non-optimal.  (5) will not be
adopted because it is slow, requiring 12 preparations of $\left| {\pi /8}%
\right\rangle _{L}$ for every Toffoli gate, and the preparation is
itself non-trivial. (6) is important because measurement of $X$,
$Y$ and $Z$ can be performed fault-tolerantly for any stabilizer
code, not just $[[n,1,d]]$ codes. Gottesman also proposed the use
of measurements and whole-block operations to swap logical qubits
between and within blocks. (7) is a nice result, but the known
fault-tolerant constructions for $T$ assume that  fault-tolerant
versions of other gates such as $^C\!\! X$ are already available, so
this `minimal' set has not so far been used to generate
fault-tolerant universal computation.

The Gottesman methods rely heavily on measurement, which might be
thought to be disadvantageous. In fact, since the measurements can
be absorbed into the recoveries (see section \ref{s:measure} and
\cite{99:SteaneB}) they are available at no cost and therefore are
advantageous. In any case all the methods involve measurement
and/or state preparation to implement the Toffoli or an equivalent
gate. Since any useful quantum computation must make significant
use of gates outside the Clifford group (otherwise it could be
efficiently simulated classically), the methods are all roughly
equivalent in this regard. For example, the speed of Shor's
algorithm to factorize integers is limited by the Toffoli gates
required to evaluate modular exponentials
\cite{96:Beckman,96:Vedral,Bk:Nielsen}.

\section{Measurement of logical Pauli observables}
\label{s:measure}

\begin{quotation}
{\bf Theorem 1.} {For any CSS code, measurement of a set $\cal M$ of
logical observables in the Pauli group can be performed at almost
no cost by merging it with a single recovery operation, as long as
the set has the following properties: every $M \in {\cal M}$ is of
the form either ${X}_u$ or ${Y}_u$ or
${Z}_u$ (i.e. a product of one type of Pauli operator),
and not all three types of operator appear in the set.}
\end{quotation}

Theorem 1 was put forward in \cite{99:SteaneB} for the case of measuring
a single observable of the form ${X}_u$, ${Y}_u$
or ${Z}_u$. The method is to prepare an ancilla in
$\ket{a_u} = \ket{0}_L + \ket{u}_L$, then operate blockwise $^C\!\! X$
or $^C \! Y$ or $^C\!\! Z$ from ancilla to data, then measure the ancilla
in the $\{\ket{+}, \ket{-}\}$ basis. The measurement outcomes
permit both an error syndrome and the eigenvalue of the relevant
observable to be deduced. The ancilla preparation is done
fault-tolerantly. One fault-tolerant method is to produce an imperfect
version of the desired state
$\ket{a_u}$ by any means, and then to measure all those observables in
the stabilizer of $\ket{a_u}$ that  consist of only $Z$ operators;
the prepared state is rejected if any of these verifying
measurements yield the wrong eigenvalue (-1), and in such cases a
further preparation attempt is initiated.
Any prepared ancilla
state that  passes the verification does not have correlated $X$
errors in it \cite{02:SteaneA}, so can safely act as the
control bits in a blockwise controlled gate with the data. $Z$
errors in the ancilla preparation (whether correlated or not)
cause the wrong syndrome and/or wrong eigenvalue of the observable
being measured on the data to be deduced. This is guarded against
by repetition and taking a majority vote. This vote corrects
the effects of $Z$ errors in the ancilla preparation; it explains
why it was not necessary to measure the $X$-type stabilizer observables in the
verification step. The whole procedure is fault-tolerant
if the noise is uncorrelated and stochastic.
It is efficient if the initial
preparation attempt has a non-negligible probability of success (i.e.
of producing $\ket{a_u}$ with no $X,Y$ or $Z$ errors).

In the method just outlined, only a subset of the observables in the stabilizer
of $\ket{a_u}$ was measured in order to verify the ancilla. Other methods are
possible. For example a measurement of the complete set of observables, combined
with rotations conditional on the outcomes, is one way to prepare $\ket{a_u}$.
Further copies could be produced and then compared by controlled-not.

To generalize to the complete result presented in the theorem,
consider first a set of observables of a single
type $\{ {M}_u \}$ where $M$ is either $X$ or $Y$ or $Z$. A
measurement of any pair ${M}_u, {M}_v$ is
equivalent, both in the eigenvalue information obtained, and in
the state projection which results, to measuring all members of
the closed Abelian group $\{I, {M}_u, {M}_v,
{M}_u {M}_v = {M}_{u + v}\}$. Similarly,
measuring the whole set is equivalent to measuring an Abelian group,
and the corresponding binary vectors $\{ u \}$ form a linear
vector space. The ancilla is prepared in
  \beq \ket{a_{\{ u \} }} = \sum_{u} \ket{u}_L
\eeq and the rest of the method proceeds as
before.

When the set $\cal M$ to be measured contains members of two
different types, the members of each type are measured during each
part of the syndrome extraction. that  is, the syndrome extraction
proceeds in two parts for CSS codes. These are normally envisaged
to collect $X$-error and then $Z$-error syndromes, but we are free
to choose any one out of the three pairs $\{X,Z\}$, $\{X,Y\}$,
$\{Y,Z\}$ to get the complete syndrome information. Each is
obtained by operating the relevant type of controlled gate from
ancilla to data, so we can simultaneously measure the same
combinations of observable types. We cannot measure single
observables of mixed type because we only have blockwise
controlled-gates of un-mixed type available.

\subsection{Logical state preparation}  \label{s:prepare}

Next we address preparation of logical states. In order to introduce
notation, let us list the simplest measurements that  theorem 1 permits,
namely measurement of $X$, $Z$ or $Y$ on any single qubit in a
block. These are indicated thus:
\begin{center}
\vspace{4 pt}
    \makebox{\includegraphics[scale=0.4]{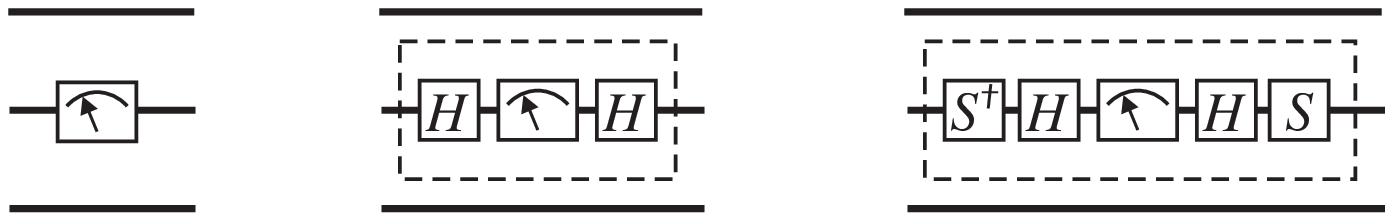}}
\vspace{4 pt}
\end{center}
Each group of lines in such a diagram represents the logical
qubits of a given block---by showing more than one we
indicate that  the operation can act on a single bit within the
block. The dotted box indicates that  the group of operations take
place in a single step.

Now, the measurement procedure is such as to leave the
encoded block in an eigenstate of the measured observable, in the
logical Hilbert space. Furthermore, it
is shown in the appendix that  we can also apply Pauli
operators to individual qubits, and groups of qubits, within a
block. It follows that  we can prepare any logical qubit in the
eigenstate of eigenvalue $+1$ of any Pauli operator (by a measurement
followed by application of an anti-commuting Pauli operator when the measured
eigenvalue is $-1$). This gives the following set of basic
fault-tolerant state preparations:
\begin{center}
\vspace{4 pt}
    \makebox{\includegraphics[scale=0.4]{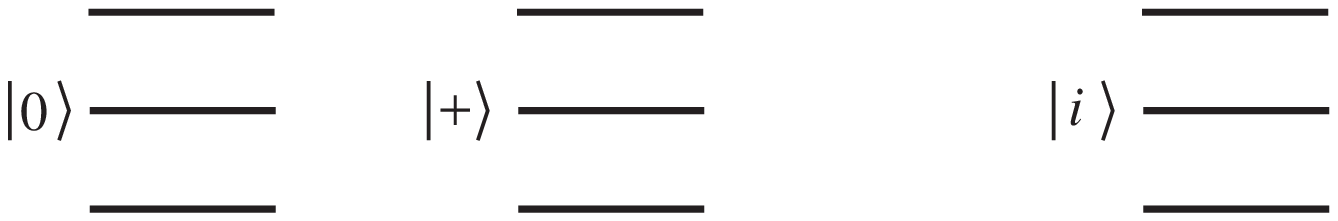}}
\vspace{4 pt}
\end{center}
where $\ket{\pm} = \ket{0} \pm \ket{1},\; \ket{\pm i} = \ket{0} \pm i \ket{1}$.

Measurements can be useful for preparing logical qubits
not only in the standard states just listed, but also in entangled states. The
class of logical states which can be prepared by the method
described is a fairly large and powerful class:

{\bf Corollary to theorem 1.} {\em Any set of logical qubits within a given block
can be prepared in a quantum codeword state of any
quantum stabilizer code whose stabilizer separates into pure-$X$
and pure-$Z$ parts, using a single recovery.}

Note, the logical qubits remain encoded in their original `inner'
code; the corollary describes the preparation of certain
superpositions of logical states. The corollary follows
immediately from the remarks above: the recoveries are used to
measure the stabilizers of the outer code, which have the right
form when the stabilizer separates as stated. The operator to move
from a $-1$ to a $+1$ eigenstate is a tensor product of Pauli
operators and so is also available.

For example, the Bell state $\ket{00}_L + \ket{11}_L$ is a quantum
codeword of a $[[2,0,2]]$ CSS code with stabilizer $XX,ZZ$. The
corollary allows us to prepare such states of pairs of logical
qubits in the same block; this is very useful for teleportation.
The following diagrams record this fact and give a slightly more
complicated example, which we will use later and which further
illustrates the method:
  \beq
    \makebox{\includegraphics[scale=0.4]{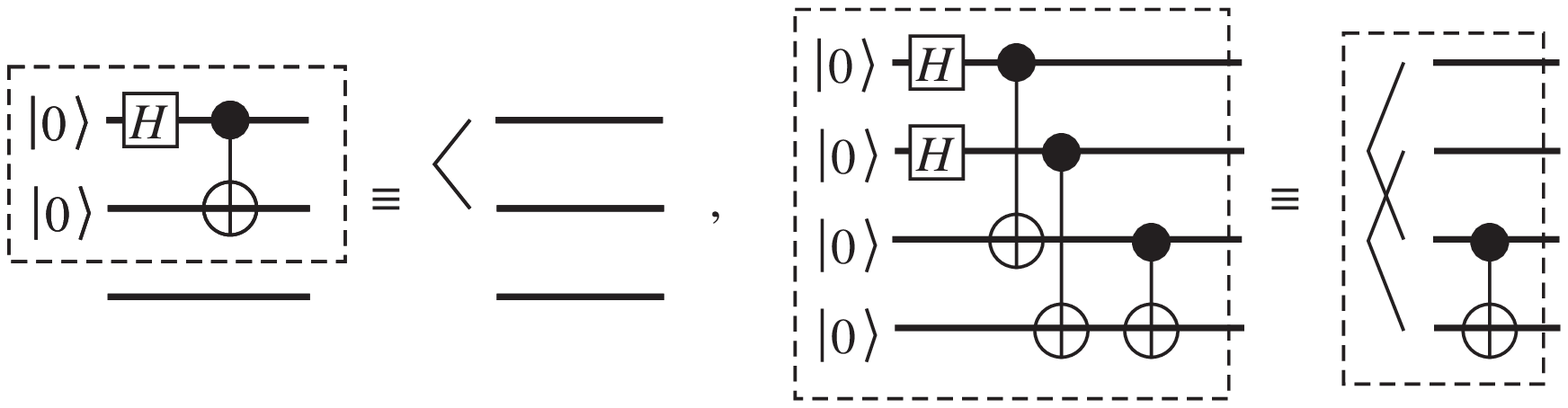}} \label{Bell}
  \eeq
The first example is used in all the constructions presented in the rest of
this paper, see (\ref{telenote}) to (\ref{Toff2}).
The stabilizer for the 2nd example is generated by $X_{1110},\;
X_{0101},\;Z_{0111},\;Z_{1010}$. For this case
the ancilla used to extract the
syndrome for $Z$ errors is prepared in $\ket{0000}_L +
\ket{1110}_L + \ket{0101}_L + \ket{1011}_L$;
the ancilla used to extract the syndrome for $X$
errors is prepared in $\ket{0000}_L + \ket{0111}_L + \ket{1010}_L +
\ket{1101}_L$.

For the sake of clarity, let us examine the ancilla preparation in a little more detail, by
using preparation of $\ket{000}_L + \ket{110}_L$ in the ancilla as an
example. Let $G_0$ and $H_0$ be the generator and check matrices
of the classical code $C_0$ which forms the zeroth quantum codeword
(see equation (\ref{cw0})). $G_0$ is $(n-k)/2 \times n$; $H_0$ is
$(n+k)/2 \times n$.

The state $\ket{000}_L$ may be prepared using a network
obtained directly from $G_0$ \cite{96:SteaneB}. To prepare $\ket{000}_L +
\ket{110}_L$ it suffices to add the single row $(110)D$ to $G_0$
and use the resulting matrix to construct the generator network
(c.f. equation (\ref{Xbar}); the expression $(110)D$ is a product
of a row vector $(110)$ with a $3 \times n$ matrix $D$).

Next we need to verify the state
against $X$ errors. The stabilizer of $\ket{000}_L +
\ket{110}_L$ has a $Z$ part consisting of $H_0$ with one row
removed, and an $X$ part consisting of $G_0$ plus the extra row
$(110)D$ (since ${X}_{110}( \ket{000}_L +
\ket{110}_L) = \ket{000}_L + \ket{110}_L$). The verification
only measures the $Z$ part of the stabilizer. To identify
the correct row of $H_0$ to remove, note that  $H_0$ consists of
the $Z$ part of the quantum code stabilizer, which has $(n-k)/2$
rows and is the same as $G_0$, plus $k$ further rows which are the
logical ${Z}$ operators. The desired state is stabilized
by ${Z}_{110}$ but not by ${Z}_{100}$
or ${Z}_{010}$. Therefore we replace the two rows
${Z}_{100}$ and
${Z}_{010}$ in $H_0$ by
the single row ${Z}_{110}$.

A useful further insight is provided by considering the quantity
of information obtained by the adapted syndrome extraction. This
can be seen from a simple counting argument, as follows. A single
quantum codeword such as $\ket{0}_L$ in a CSS code is an equal
superposition of $2^{\kappa}$ product states in the computational
basis, where $\kappa = (n-k)/2$ is the size of the classical code
$C_0$ (equation (\ref{uL})). The Hadamard transformed state is
then an equal superposition of $2^{n - \kappa}$ product states.
When we are using such a state to extract an error syndrome, for a
zero syndrome we expect to observe one of these $2^{n-\kappa}$
states. Correctable errors will transform the state onto an
orthogonal one. There is a total of $\kappa$ bits of remaining
room in Hilbert space for mutually orthogonal sub-spaces, so the
measurement yields $\kappa$ bits of information, this is the error
syndrome (for either $X$ or $Y$ or $Z$ errors). If instead the
state was originally prepared in $\ket{0}_L + \ket{u}_L$, then it
consisted of an equal superposition of $2^{\kappa + 1}$ product
states. Upon being Hadamard transformed, it becomes an equal
superposition of $2^{n-(\kappa+1)}$ states, hence there are
$\kappa + 1$ bits of information about what has happened to it
available from measurements on it. These are the error syndrome
and the eigenvalue of the measured observable, which are commuting
observables so can be simultaneously measured. The argument
extends in an obvious manner when further mutually commuting
observables are measured.

\subsection{More general state preparations} \label{s:general}

The available tools for state preparation can be extended
as follows. We wish to prepare a state $\ket{\phi}_L$ of $k$ logical
qubits that  is
uniquely specified by a set $\{ M_i \},\;(i=1 \cdots k)$ of $k$ linearly
independent commuting
observables; this set generates the stabilizer of $\ket{\phi}_L$
in the logical Hilbert space.
If $\ket{\phi}_L = G \ket{0^{\otimes k}}_L$ then one possible
choice of the stabilizer operators is \cite{00:Zhou} ${M}_i = {G} {Z}^i
{G}^{\dagger}$. Define ${Q}_i = {G} {X}^i {G}^{\dagger}$, then
each $Q_i$ anticommutes with its associated stabilizer operator
and commutes with all the others: $M_i Q_i = - Q_i M_i$ and $[M_i,
Q_{j \ne i}] = 0$. The $M_i$ and the $Q_i$ all have eigenvalues
$\pm 1$.

One method to prepare $\ket{\phi}_L$ is to measure all the $M_i$ on
some arbitrary input state in the code space, and
whenever an eigenvalue $-1$ is found, apply the operator $Q_i$
that  moves the $-1$ eigenstate to the $+1$ eigenstate.
However, it may not be straightforward to measure one of more of
the $M_i$ fault-tolerantly.

Let $M_r$ be a stabilizer operator
whose fault-tolerant measurement is not straightforward.
Decompose it as $M_r = N_{r,1} \otimes N_{r,2} \cdots \otimes N_{r,p}$ where
there exists a state which is a $+1$ eigenstate of all the $N_{r,j}$ simultaneously,
and where the $N_{r,j}$
are simpler to work with fault-tolerantly than $M_r$, for example
because they each act on fewer qubits. To prepare $\ket{\phi}_L$, first
prepare a $+1$ eigenstate of all the
$N_{r,j}, (j=1 \cdots p)$ (e.g. by measuring them if they commute), and then measure all the
other $M_{i \ne r}$. Typically the $N_{r,j}$ will not commute with
all the $M_{i \ne r}$, but as long as the measurements are done in
the order described the final state is the same as if $M_r$ had
been measured.

For example, suppose we require the input state
  \beq \ket{\phi}_L = \ket{00}_L - \ket{11}_L + \ket{01}_L + \ket{10}_L . \eeq
This has stabilizer $X_{10}Z_{01} = {XZ}$, $X_{01}Z_{10} = {ZX}$.
Neither of these observables can be measured easily, but the
product $(XZ) \, (ZX) = {YY}$ can, since it is not of mixed type.
We therefore adopt the set $\{M_i \} = \{ XZ,\; YY \}$.
Decomposing $M_1 = XI \otimes IZ$, we see it is sufficient to prepare a
$+1$ eigenstate of $X$ in the first qubit, and of $Z$ in the
second qubit, which is easy: the starting state is $\ket{00}_L +
\ket{10}_L$. Upon measuring $YY$ (and applying $IZ$ if the
measured eigenvalue is $-1$), $\ket{\phi}_L$ is obtained.

It was pointed out in \cite{00:Zhou} that the starting state which
will produce $\ket{\phi}_L$ when a single stabilizer observable $M_i$ is
measured is the state $(I+Q_i) \ket{\phi}_L$. This observation can also
help in identifying suitable starting states.

We can go further and split up further $M_i$
operators into their components $N_{i,j}$ as long as
a +1 eigenstate of all the $N$ operators at once can be
prepared. For example, the state required for the Toffoli gate discussed
in section \ref{s:Toff} has a set of 8 stabilizer generators including
$X^1 X^5 \, ^C\!\! X^{67}$, $X^2 X^6 \, ^C\!\! X^{57}$, $Z^1 Z^5$ and
$Z^2 Z^6$. We split the first two of these into $X^1
X^5$ and $^C\!\! X^{67}$, $X^2 X^6$ and $^C\!\! X^{57}$ respectively.
Preparing the 7th bit in $\ket{+}_L$ is sufficient to ensure a $+1$
eigenstate of both the controlled-gates. At the same time we
prepare the 1st and 5th bits in the Bell state $\ket{00}_L
+ \ket{11}_L$ to ensure they are in a +1 eigenstate of
$X^1 X^5$ and $Z^1 Z^5$, and similarly for the 2nd and 6th bits---
see (\ref{Toff2}).

\section{A fault-tolerant toolbox}  \label{s:toolbox}

We will now summarize some basic fault-tolerant operations and
methods that will be used in the constructions to be described.

We restrict attention to CSS codes based on a doubly-even
classical code that is contained by its dual. For such codes the
following fault tolerant operations are easily available (see
appendix):

\begin{center}
\vspace{4 pt}
    \makebox{\includegraphics[scale=0.5]{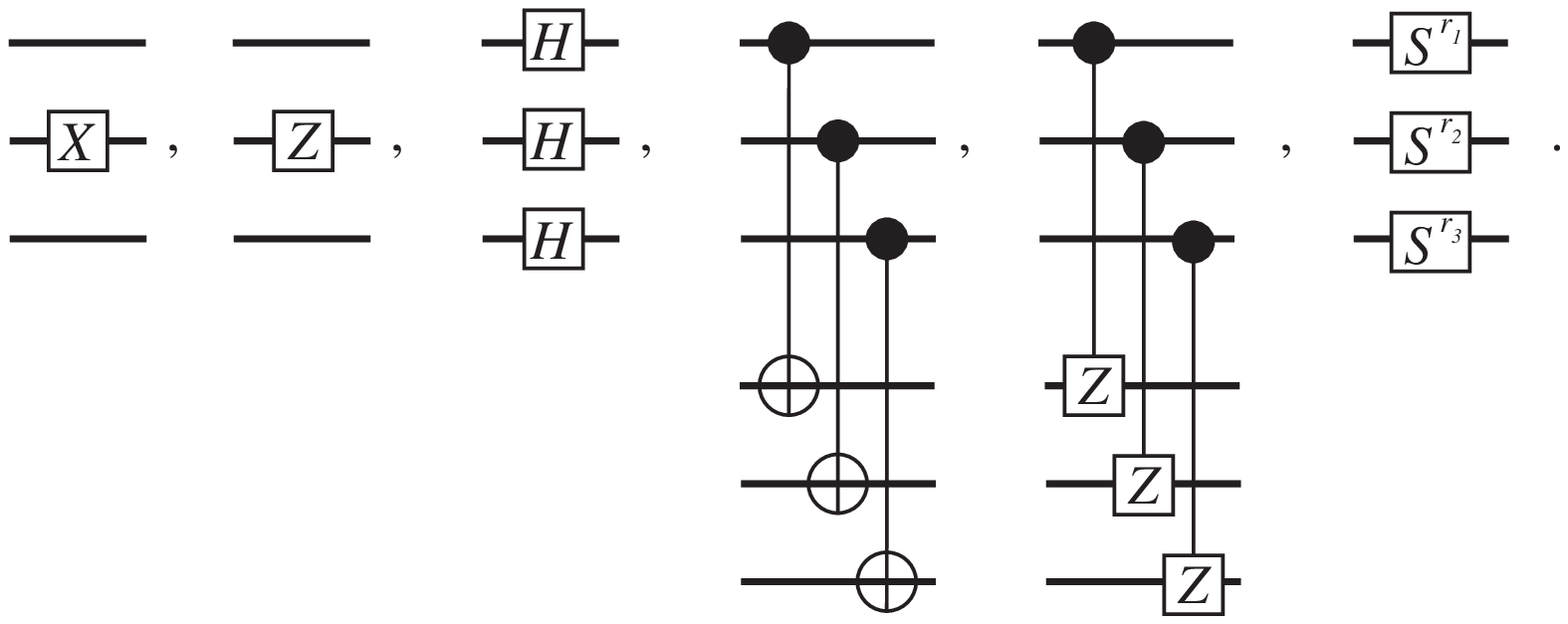}}
\vspace{4 pt}
\end{center}

\begin{enumerate}
\item Operators in the Pauli group, acting on any logical qubit or group
of qubits in a block.
\item Blockwise ${H}$ and  $^C\!\! {X}$ and hence
$^C\!\! {Z}$.
\item ${S}$ acting blockwise but such that different logical
qubits may be acted on by different powers of $S$, depending on
the code (see lemma 4 in appendix).
\end{enumerate}

\subsection{Transfer operation}

Gottesman \cite{98:GottesmanA} introduced the operation by which a state is
transferred from one qubit to another by a single $^C\!\! X$ gate and
a measurement, and its use in stabilizer codes to move a single
qubit between blocks:
  \beq
    \makebox{\includegraphics[scale=0.4]{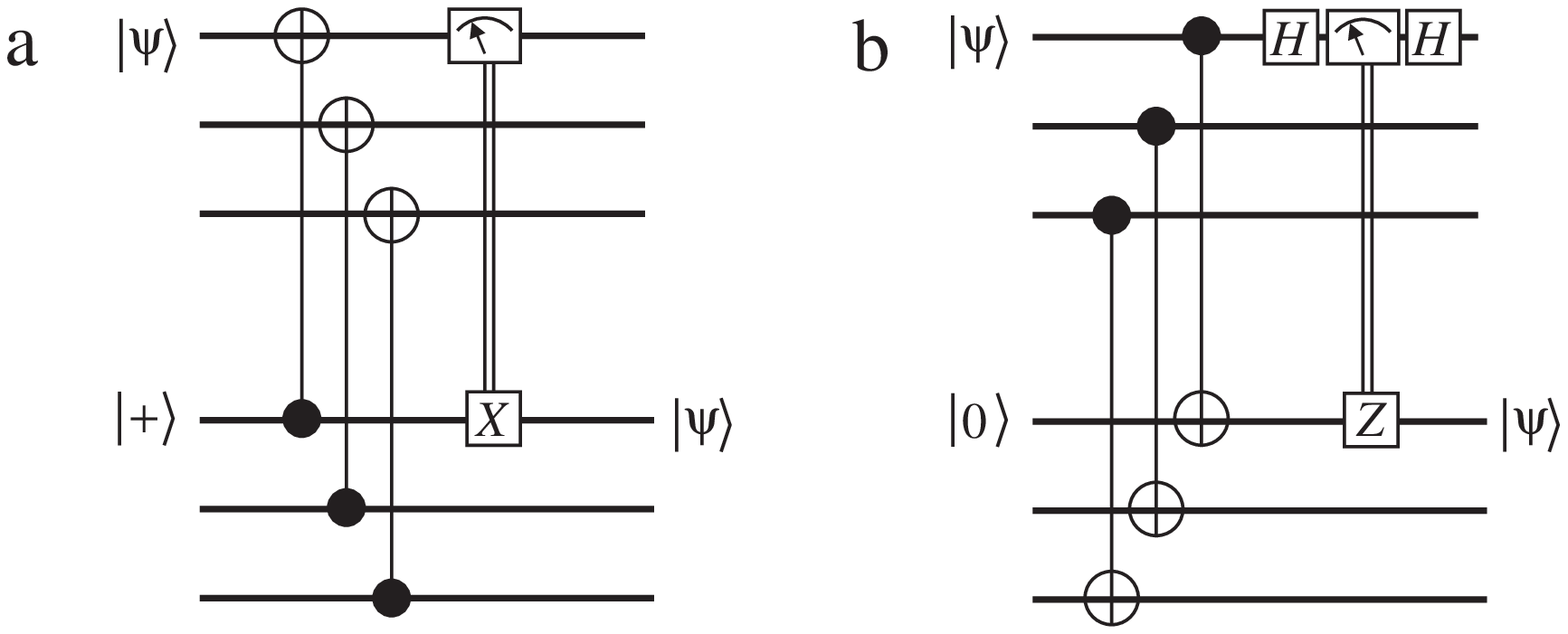}} \label{transfer}
\eeq
(\ref{transfer}) shows two versions of the operation (referred to as examples
of `one-bit teleportation' in \cite{00:Zhou}).
Since $^C\!\! X$ acts as an identity operator
when either the control bit is in $\ket{0}$ or
the target in $\ket{+}$, we can ensure the blockwise $^C\!\! X$
does not disturb other qubits in either the source block or
the destination block, by preparing states accordingly. The
next set of diagrams introduce a shorthand notation for transfer
operations of the first type in (\ref{transfer}), illustrating
various possibilities for the state preparations. In the first case
a qubit is transferred out of a full block without disturbing the
other bits in that block; in the last case a qubit is transferred into
a full block without disturbing the other bits there; the middle example
is an intermediate case:
  \beq
    \makebox{\includegraphics[scale=0.4]{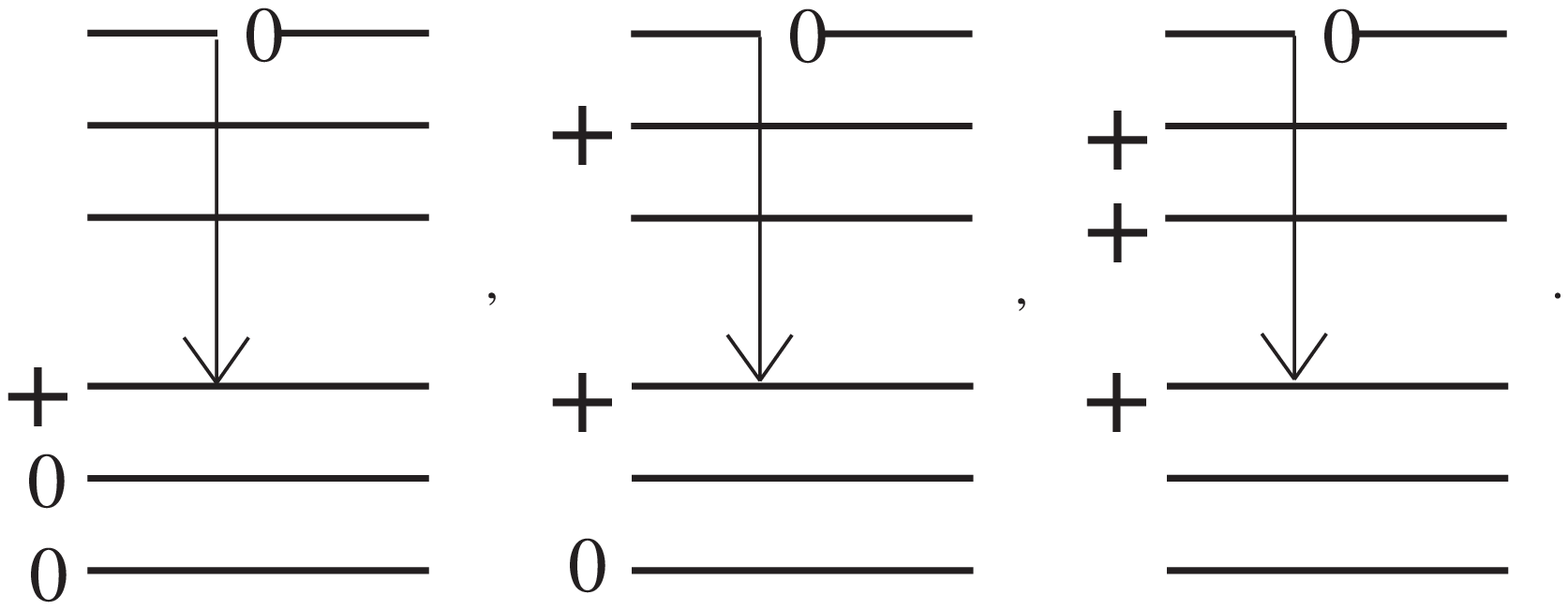}}
\eeq
The broken line followed by a zero is shorthand for measurement in
the $\ket{0},\, \ket{1}$ basis followed by $X$ if the $-1$ eigenvalue
was obtained, thus leaving the qubit in state $\ket{0}$. The relevant
point is that this state preparation does not need a further recovery,
so it takes place in the same time-step as the rest of the transfer operation.

An illustrative set of possible transfer operations of the second
type in (\ref{transfer}) is:
  \beq
    \makebox{\includegraphics[scale=0.4]{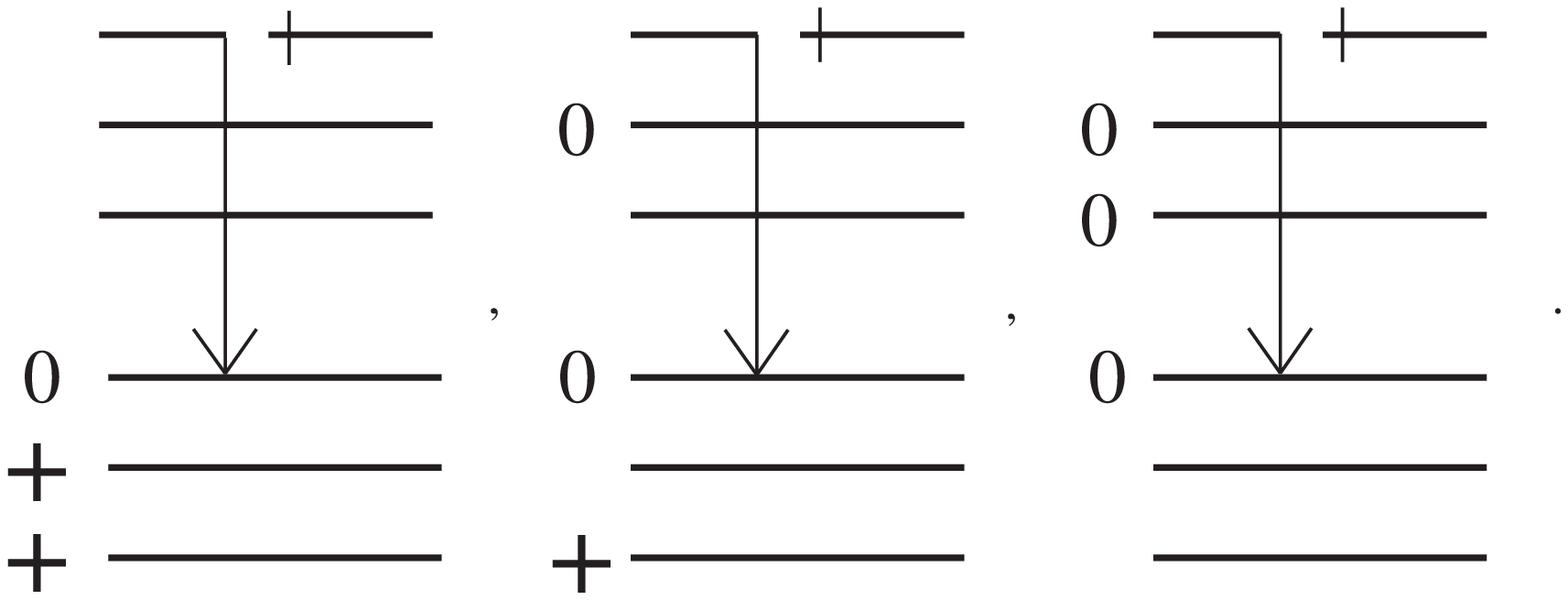}}
\eeq
The vertical bar after the line break is shorthand for preparation of
$\ket{+}$, that takes place via the measurement in (\ref{transfer}).

\subsection{Teleportation}

We define the following notation for teleportation:
\beq
    \makebox{\includegraphics[scale=0.4]{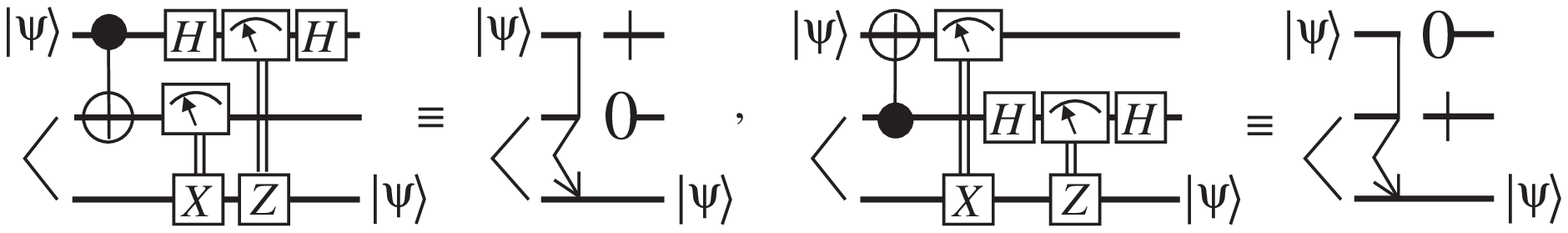}}
    \label{telenote}
\eeq
This is used to move a qubit from one block to a different location in another
block:
\beq
    \makebox{\includegraphics[scale=0.4]{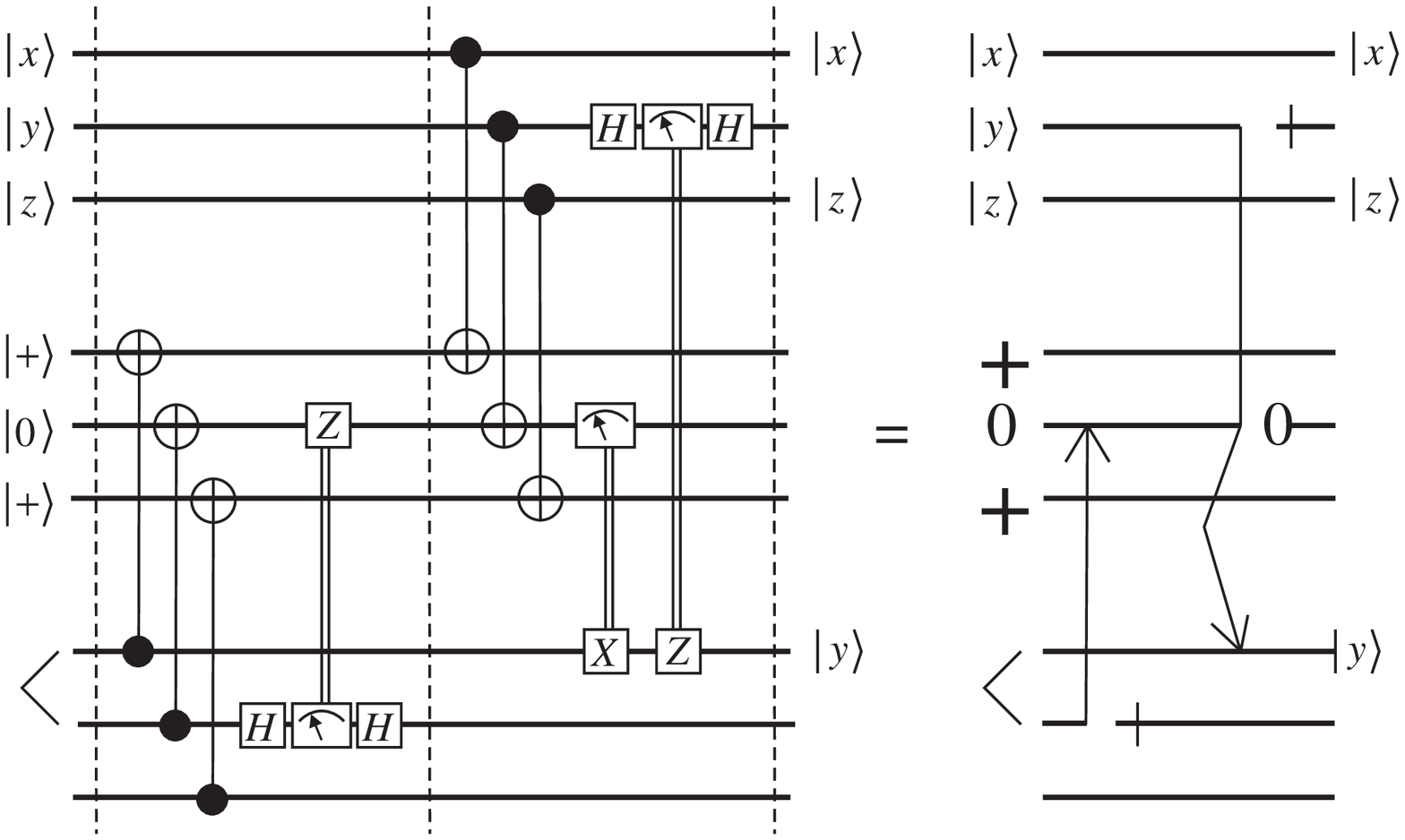}} \label{tele1}
\eeq
The initial Bell state preparation is done by a single recovery as in (\ref{Bell}),
so the complete network requires 3 time steps, these are shown separated by dashed
vertical lines.

The qubit is moved from the $i$'th position in the source block to the $j$'th
position in the destination block. The network construction is straightforward
when both the $i$'th and $j$'th qubits of the destination block are available
to be prepared in the Bell state, as in (\ref{tele1}).
The next network shows how to accomplish teleportation from a full block to
another which has only one unused position. This requires two transfers
to put the Bell state in the right place, and a naive construction would require
4 time steps. However, the second transfer can take place simultaneously with
the teleportation step:
\beq
    \makebox{\includegraphics[scale=0.4]{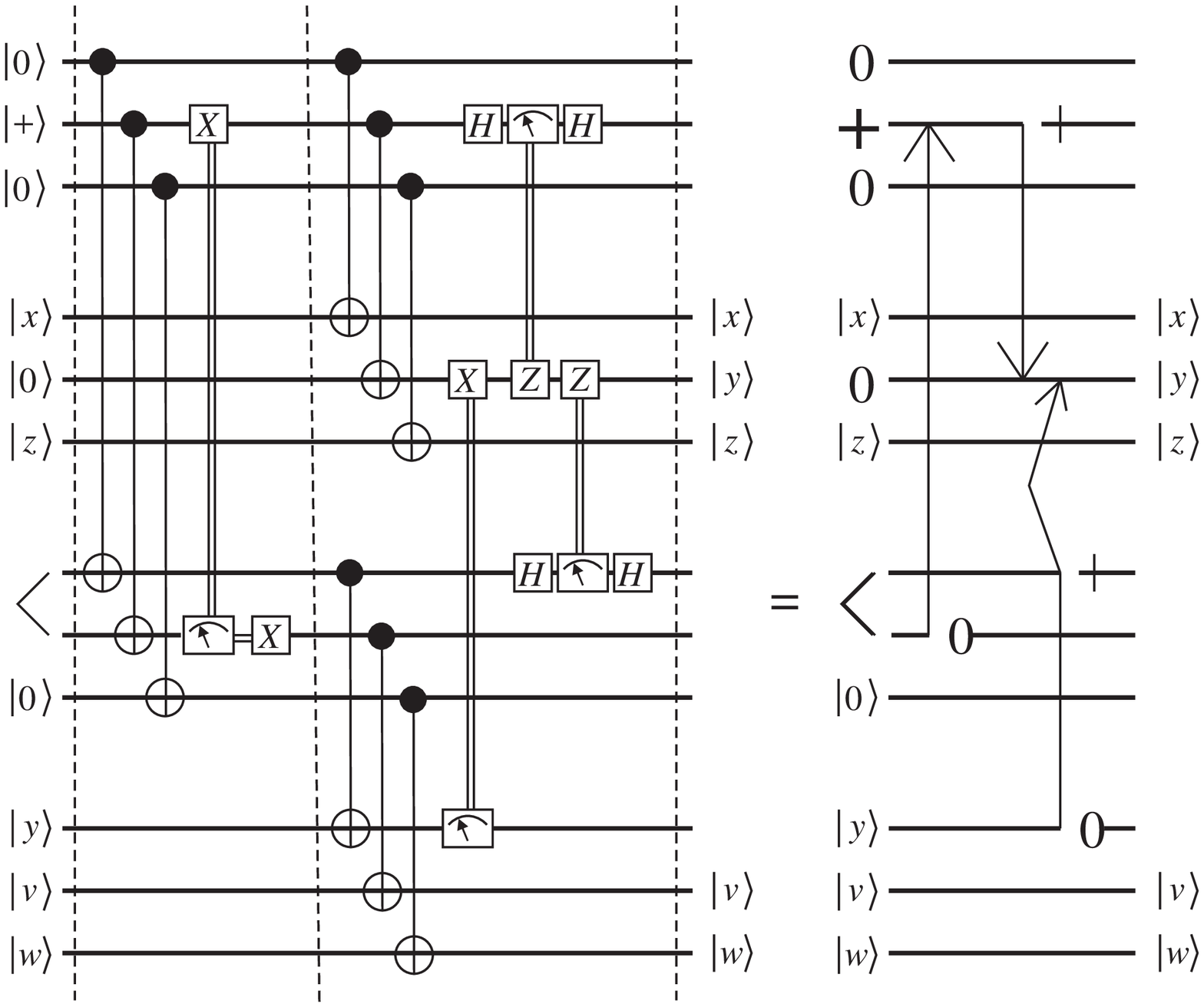}}  \label{back}
\eeq
This network has the interesting feature that the transfer and teleportation
in the final step commute, and therefore are applied simultaneously. One way
to `read' the network is to argue that the upper of the two simultaneous
blockwise $^C\!\! X$ gates creates a GHZ state $\ket{000} + \ket{111}$ between
the middle bits of the 1st two blocks and the upper bit of the 3rd; this
entangled triplet replaces the entangled pair in the standard teleportation.
$X$-measurements on two of these qubits are then needed to disentangle
them from the one which is teleported.

\section{Controlled-not}  \label{s:cnot}

We now turn to implementing $^C\!\! X$ between any single pair of qubits. We
treat the case where the qubits are in the same block, which will illustrate
all the essential ideas.

One method is to use two teleportations and a blockwise $^C\!\! X$. A naive
construction would require $3+1+3=7$ time-steps, but by choosing
transfer operations that leave states ready-prepared for the
subsequent step, and combining steps where possible, this is
reduced to 5: \beq
    \makebox{\includegraphics[scale=0.4]{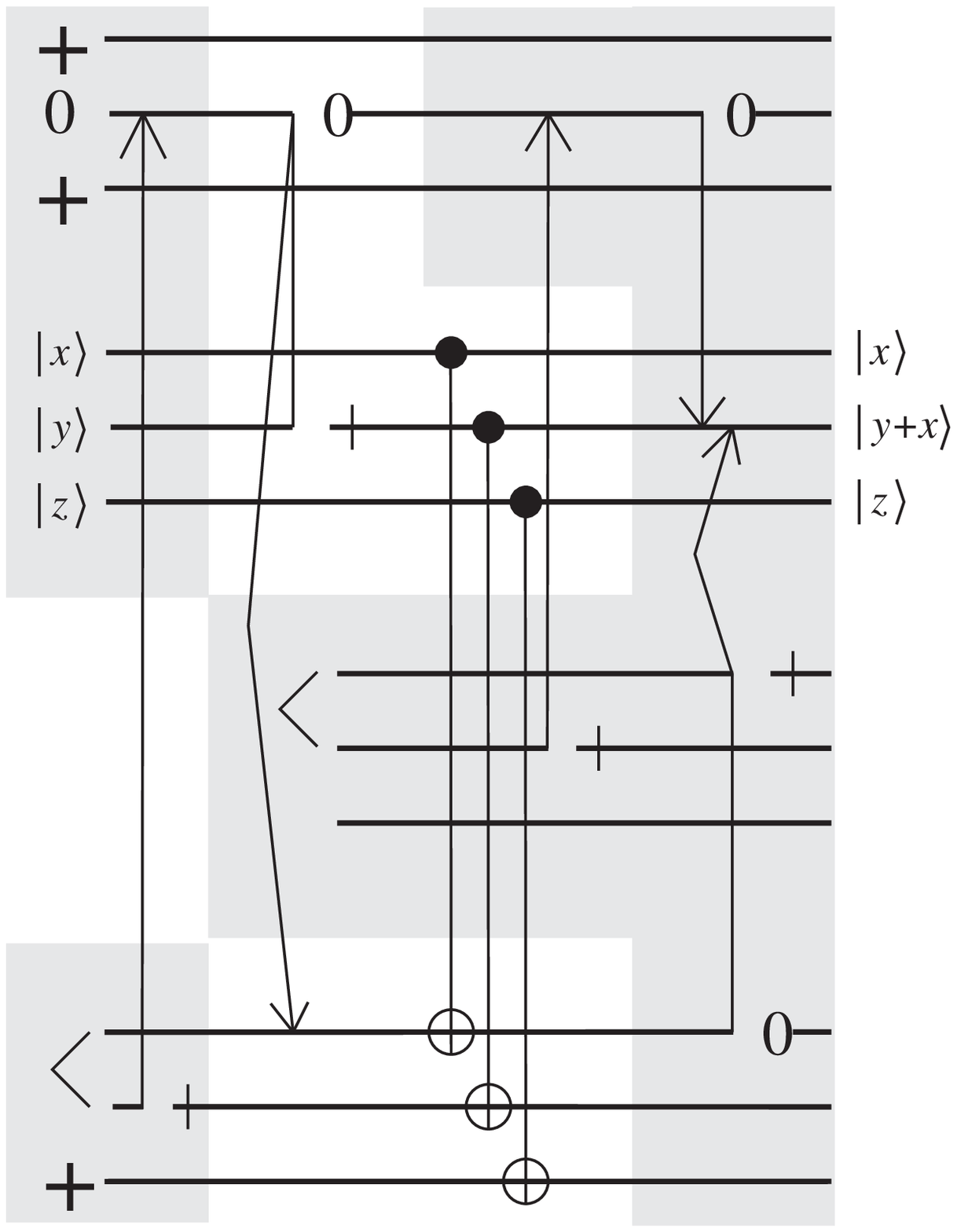}}   \label{cnota}
\eeq
The shaded area is the offline part, where, as discussed in section \ref{s:resource},
we count the initial teleportation (from `memory' to `accumulator')
as online, and the final teleportation (from `accumulator' to `memory')
as offline.

The Bell-state measurement that forms part of the standard
teleportation operation, see (\ref{telenote}), begins with a $^C\!\!
X$ gate involving one of the qubits of the entangled pair.
However, when using whole-block operations it is easier to
implement a group of $^C\!\! X$ gates such that both qubits of the
entangled pair are operated on (either as target or control bits).
We therefore consider the following network which teleports the
second logical qubit (initially in state $\ket{y}_L$), where the
initial blockwise $^C\!\! X$ is implemented without insisting that the
first qubit is prepared in $\ket{0}_L$ (it is in some general
state $\ket{x}_L$ instead):
  \beq
  \makebox{\includegraphics[scale=0.4]{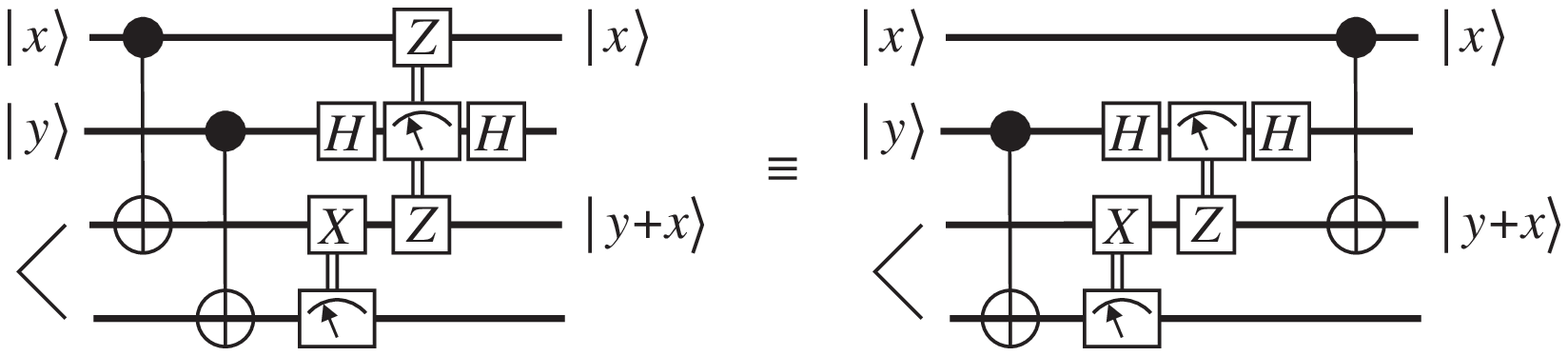}}  \label{cnotb}
  \eeq
This shows that the result is a $^C\!\! X$ operation between the first
and second bits, with the second output bit teleported into the second
block. Equation (\ref{cnotb}) may be derived by starting with the
right hand side (which shows a teleport followed by $^C\!\! X$) and
commuting the final $^C\!\! X$ backwards, as in
\cite{99:GottesmanB,00:Zhou}. To complete the $^C\!\! X$ operation,
the target qubit can be teleported back to its original block at
the end as in (\ref{back}). Using similar ideas to those in
(\ref{cnota}), the complete network, including gathering the
qubits into one block at the end, requires 4 blocks and 3 time
steps, of which 1 is online.

The concept behind equation (\ref{cnotb}) can be extended
so as to achieve networks
of ${\cal C}_2$ gates involving up to half the qubits in a block in
a single online step, as long as the network finishes with a set of
$^C\!\! X$ gates connecting the non-teleported bits to the teleported ones.
For example:
\beq
    \makebox{\includegraphics[scale=0.4]{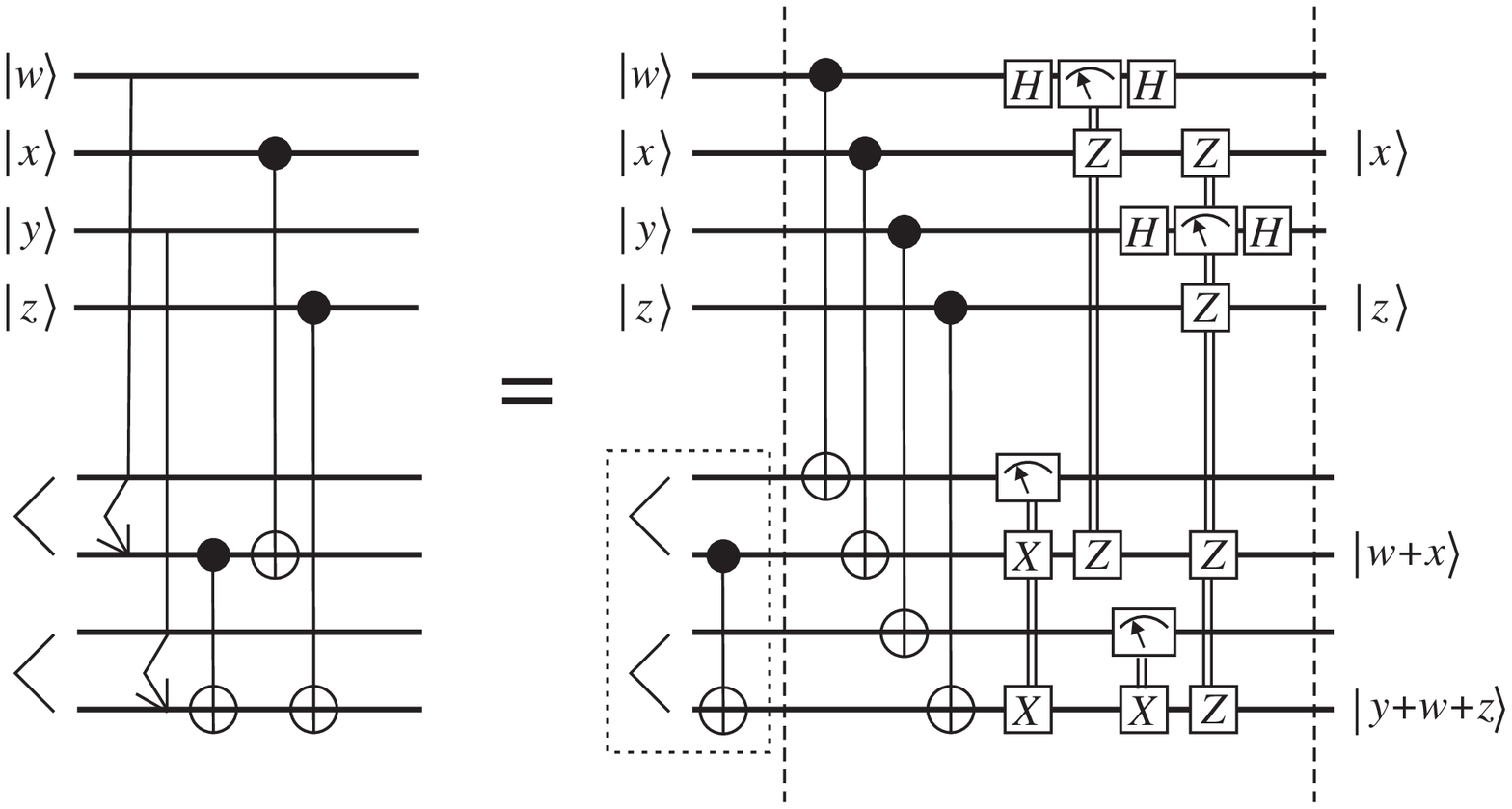}}  \label{cnotb3}
\eeq
The initial preparation step is an example of the corollary to theorem 1.
This network is an example of a class of networks discussed below in
connection with theorem 2.

When $x=z=0$, (\ref{cnotb3}) is an example of the general method
introduced by Gottesman and Chuang in \cite{99:GottesmanB}.

If we introduce a further ancilliary block, the Gottesman-Chuang
method can achieve $^C\!\! X$
between bits in the same block while teleporting the whole block,
thus keeping its constituent logical bits together:
\beq
    \makebox{\includegraphics[scale=0.4]{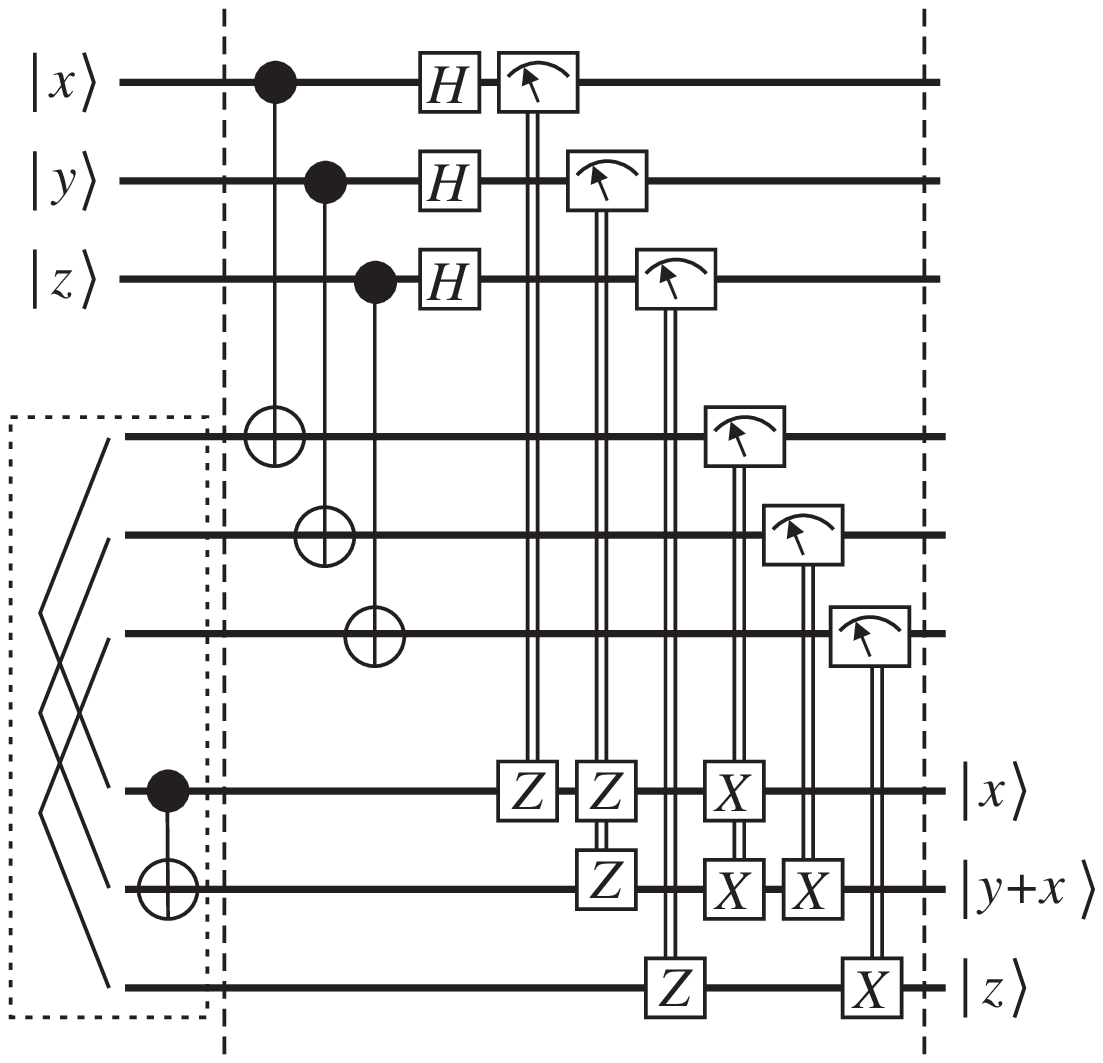}} \label{cnotc1}
\eeq
The offline state preparation shown in the dashed box can be accomplished
in three time steps, by making use of the following equivalences:
\beq
    \makebox{\includegraphics[scale=0.4]{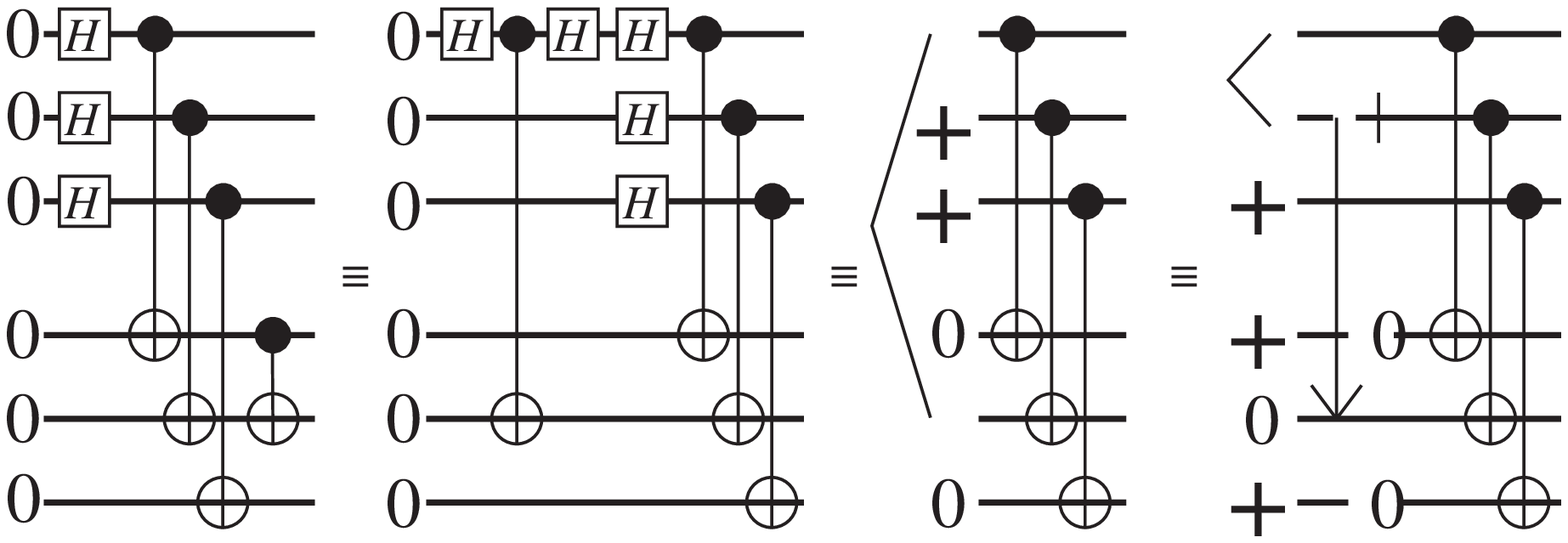}}   \label{cnotc2}
\eeq
The zeros just after the transfer operation represent state preparations that  
take place at the same time as the transfer. They ensure the final blockwise
$^C\!\! X$ in (\ref{cnotc2}) has the correct entangling effect.

The resources required by the $^C\!\! X$ constructions of equations
(\ref{cnota}), (\ref{cnotb}), (\ref{cnotc1}) are summarized in table 1.

\begin{table}
\begin{tabular}{c|cc|cc|cc}
 & \multicolumn{2}{c}{Network (\ref{cnota})} & \multicolumn{2}{c}{(\ref{cnotb})}
& \multicolumn{2}{c}{(\ref{cnotc1})}  \\
& offline & online & off. & on. & off. & on. \\
\hline
blocks     &  4 & 4   & 4 & 2   & 2 & 3 \\
time steps &  3 & 2   & 2 & 1   & 3 & 1 \\
area       & 13 & 5   & 9 & 2   & 6 & 3
\end{tabular}
\caption{Summary of resources required by three networks for $^C\!\! X$ between
bits in the same block.}
\end{table}

\subsection{Discussion}

For a code with $k=1$ the gate we have discussed would be trivial: a
single transversal $^C\!\! X$ suffices, followed by a single recovery.
It is noteworthy that the more complicated (but more
space-efficient) codes with $k>1$ can achieve the gate without any slow-down:
the online parts of (\ref{cnotb}) and (\ref{cnotc1}) require only a single
time step. Similar constructions can be found for other operators in the group
${\cal C}_2$, using the general insight of commuting gates backwards through
teleportations \cite{99:GottesmanB,00:Zhou}. The main contributions of the present study
are the extended use of recovery operations for preparing entangled states (avoiding the need for
cat states), the minimization of time steps by careful construction in (\ref{back}),
(\ref{cnota}), (\ref{cnotc2}), and the possibility of multi-qubit networks of ${\cal C}_2$ gates
in a single online step, as illustrated by (\ref{cnotb3}). We now generalize the latter point.

\begin{quotation}
{\bf Theorem 2.} {Any network of gates in ${\cal C}_2$ (the Clifford group)
can be applied fault-tolerantly to any group of logical bits (in the same or different
blocks) using a single online time step.}
\end{quotation}

{\em Proof:} The result is obtained from applying the
Gottesman-Chuang method illustrated in (\ref{cnotc1})
not just to single gates such as $^C\!\! X$ or $H$, but to
networks of gates.
Suppose the bits involved in the network occupy $N$ blocks. They
are all teleported using $N$ pairs of blocks.
As long as all the gates in the network to be implemented are in ${\cal C}_2$,
they can all be commuted backwards through the Pauli operations involved in the
teleportations such that still only Pauli operations are required
to complete the teleportation. The final Pauli operations can then be applied all at
once immediately after the measurements.

Diagram (\ref{cnotb3}) illustrates a related result: some networks
of ${\cal C}_2$ gates can be implemented among bits in a single block using
only a single extra block.

\section{Toffoli gate}  \label{s:Toff}

Following \cite{99:GottesmanB}, we will use the following type of construction for the Toffoli gate:
\beq
    \makebox{\includegraphics[scale=0.4]{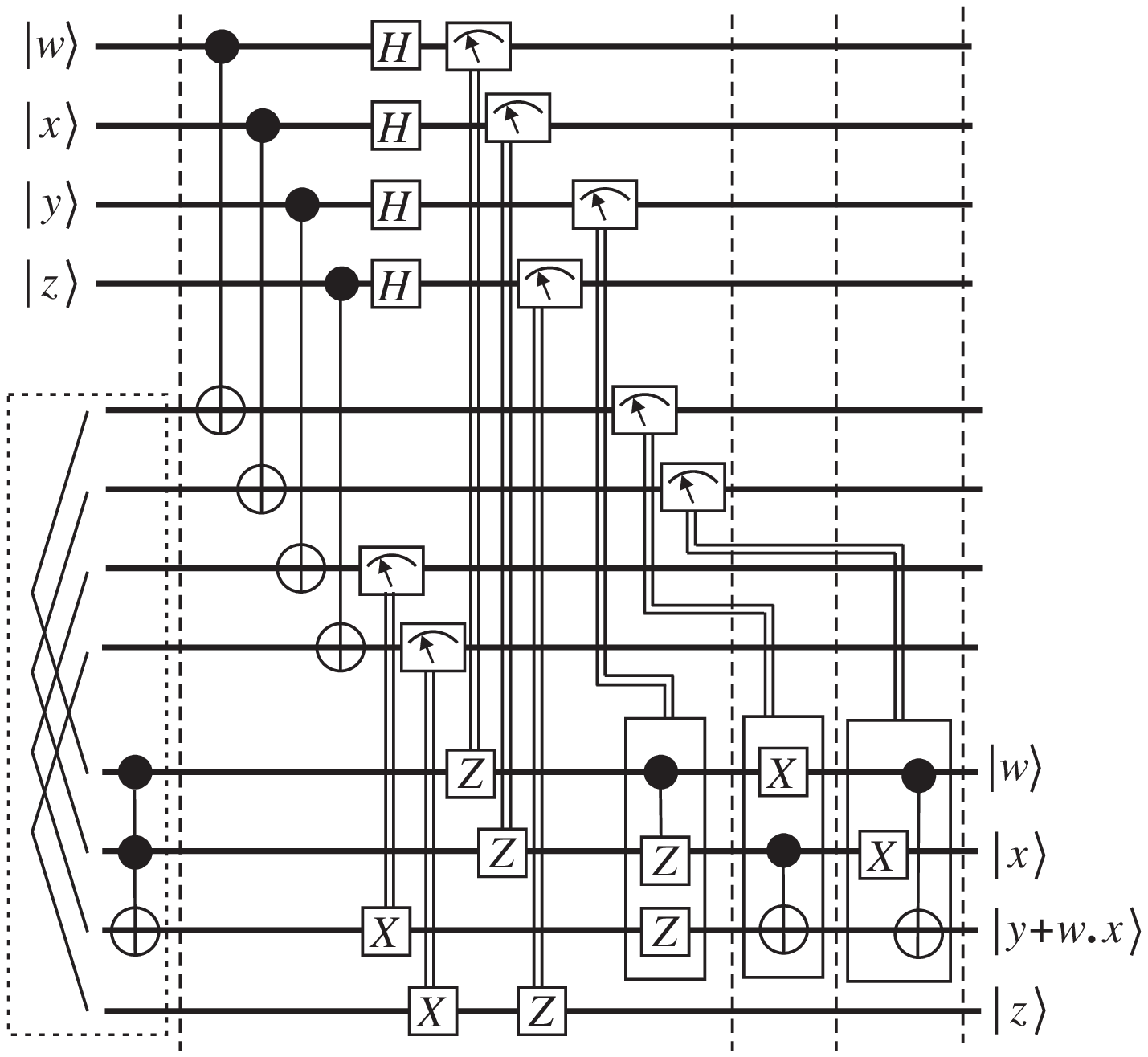}}   \label{Toff}
\eeq
This approach, rather than Shor's original network
(related to one-bit teleportation, see \cite{00:Zhou}) is adopted because it lends itself
better to blockwise operations. In (\ref{Toff}) a fourth qubit of each block is
included in order to show what happens to the rest of the bits
that are not involved in the gate itself.

In order to keep the network as rapid as possible, all the measurements should take place
together, and then whichever of the further operations are needed (conditional on the
measurement results) should be applied as soon as possible. This flexibility in timing
of the final operations is not shown in the diagram.

The dashed box is an offline preparation which we will discuss below.
Of the 8 measurements in (\ref{Toff}), 5 involve single-bit operators that can be
applied (when needed) in the same time step as the blockwise $^C\!\! X$ and the measurements
themselves. The other three involve 2-bit gates. Using the methods of either (\ref{cnotb})
or (\ref{cnotc1}) each such gate needs only a single online time step, as long as sufficient
spare blocks are available for offline preparations and/or teleportations. However, they
cannot all take place simultaneously if we retain the condition that only one
two-block gate involving any given block is allowed per recovery, to prevent avalanches of errors.
Of the 8 equiprobable measurement outcomes of this group
of 3 measurements, one requires no action, three require a single time-step,
three require 2 time-steps and one requires 3. The average number
of online time steps required by the complete network is therefore $13/8 \simeq 1.6$.

\begin{table}
\begin{tabular}{|c|c|}
\hline
\rule{0 pt}{13 pt} $M_1 = X^1 X^5 \,^C\!\! X^{67}$ & $Q_1 = Z^1$ \\
$M_2 = X^2 X^6 \,^C\!\! X^{57}$ & $Q_2 = Z^2$ \\
$M_3 = X^3 X^7$  & $Q_3 = Z^3$ \\
$M_4 = X^4 X^8$  & $Q_4 = Z^4$ \\
$M_5 = Z^1 Z^5$  & $Q_5 = X^1$ \\
$M_6 = Z^2 Z^6$  & $Q_6 = X^2$ \\
$M_7 = Z^3 Z^7 \,^C\!\! Z^{56}$  & $Q_7 = X^3$ \\
$M_8 = Z^4 Z^8$  & $Q_8 = X^4$\\
\hline
\end{tabular}
\caption{Stabilizer operators $M_i$ for the input state in the Toffoli gate network,
with their associated anticommuting operators $Q_i$.}
\end{table}

Let $\ket{\phi}_L$ be the state we need to prepare, as defined by the dashed box in (\ref{Toff}).
The stabilizer of $\ket{\phi}_L$ is generated by the operators
listed in table 2. Five of these operators are in the Pauli group ${\cal C}_1$, three are
not in ${\cal C}_1$ but are in the Clifford group ${\cal C}_2$. Fault-tolerant measurement
of the 5 Pauli group operators can be done through a recovery as in section \ref{s:prepare}.
Fault-tolerant measurement of the 3 Clifford group operators can be done by Shor's cat state
method \cite{96:Shor}. Shor described the method as applied to certain $[[n,1,d]]$ CSS codes, we
generalize it in the appendix to $[[n,k,d]]$ codes of the type under discussion (lemma 5).

We would like to minimise the need to prepare cat states. Recalling the discussion
in section \ref{s:general}, we can factorize the stabilizer operators in any convenient way
and prepare a $+1$ eigenstate of the component operators $N_{r,j}$. By this means
it is possible to avoid the need to measure any two out of
$M_1$, $M_2$ and $M_7$. For example, the discussion at the end of section
\ref{s:general} showed how to avoid the need to measure $M_1$ and $M_2$. The complete
state preparation indicated by the dashed box in (\ref{Toff}) is then obtained with
\beq
    \makebox{\includegraphics[scale=0.4]{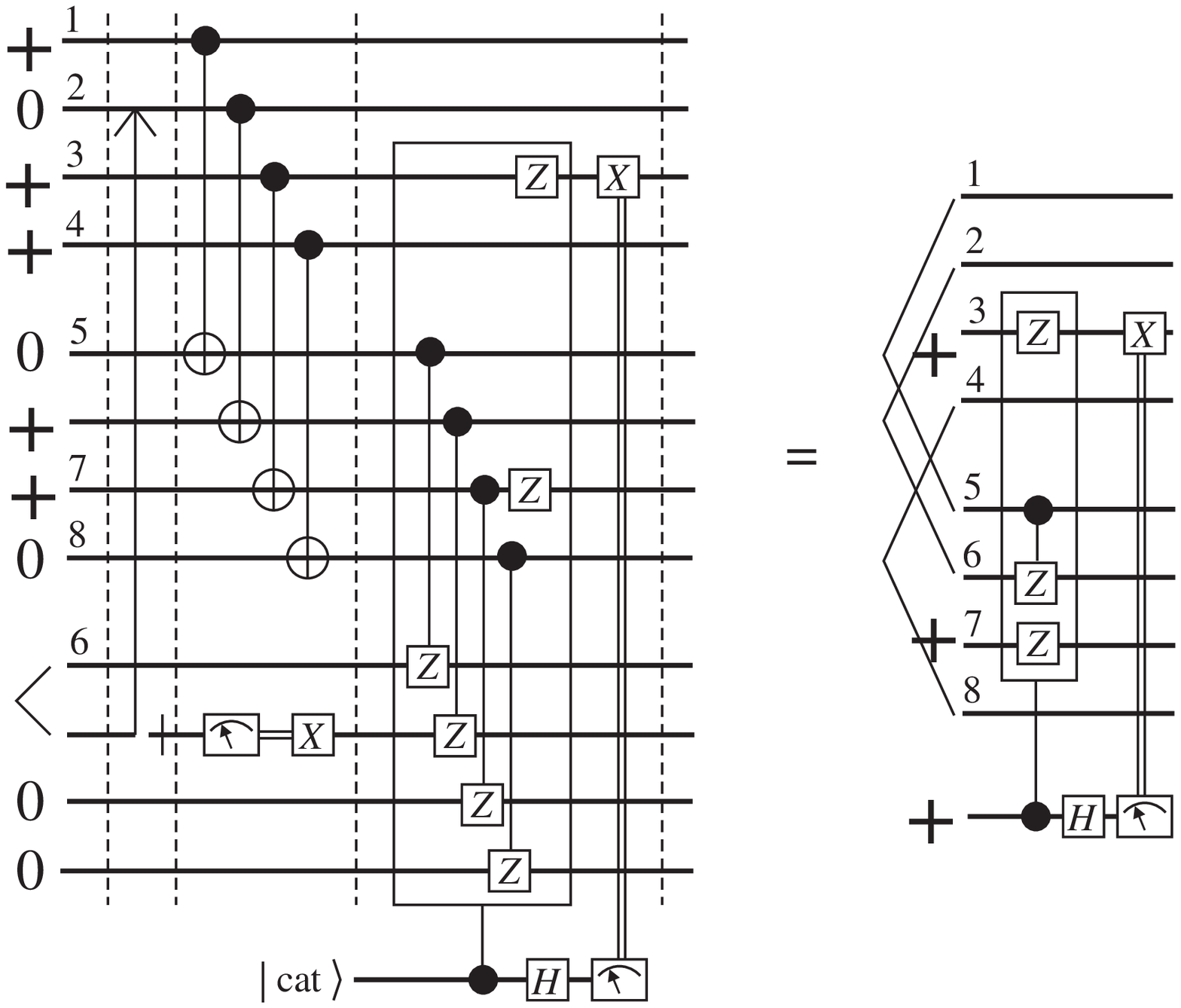}}   \label{Toff2}
\eeq
where the diagram on the right explains the logical effect of the fault-tolerant diagram
on the left. Bit number 6 has been left in a separate block so that if the $^C\!\! Z^{56}$
gate in (\ref{Toff}) is needed then it can be implemented immediately. To minimise the number
of online time steps bit 7 should also be positioned in a separate block. This can be done
using the same Bell-state preparation followed by transfer as is indicated in (\ref{Toff2})
for bits 2 and 6. The diagram shows an alternative approach that uses fewer blocks. Bit
6 (and 7 if necessary) can be repositioned back into the same block as 5 and 8 by teleportations
after the end of (\ref{Toff}).

\subsection{Discussion}

The network for the Toffoli gate between bits within a block involves at least 5 blocks
(one of which is used for the cat state) and
1 cat-state-based measurement.
The average number of online time steps is $13/8$
if a further block is used, and slightly more than this otherwise. The main result is
that the number of online time steps is independent of $k$, and in particular is
the same for $[[n,k,d]]$ codes with $k > 1$ as for $k=1$. Similar methods apply
to other gates in the class ${\cal C}_3$.

\section{Conclusion}  \label{s:conc}

We have considered fault-tolerant networks for
logic operations on bits encoded in
CSS codes, concentrating on codes based on a doubly-even classical code that is
contained by its dual (some of the methods are more general).
We have shown how to extend the use of the recovery operation to allow
preparation of an interesting class of logical states (theorem 1 and its corollary).
The implementation of certain networks in a single online time step
(theorem 2) is implicit in the Gottesman-Chuang work; we have shown
that the offline state preparation for such networks
can be accomplished efficiently using theorem 1.

We have presented optimized constructions of fault-tolerant
networks for all the members of a universal set of operations. The optimization is
primarily to minimise on-line time steps, where
one `time step' is defined to include a single recovery of the whole computer.
The constructions show that fault-tolerant operations for $[[n,k>1,d]]$ codes
require the same number of time steps as those for $[[n,1,d]]$ codes.
It follows that the total number of recoveries needed
to implement a complete algorithm is the same when $k>1$ as
when $k=1$. The number of individual block recoveries is smaller when
$k>1$ because then there are fewer blocks, assuming the computer has
more memory blocks than workspace.

We would like to thank D. Lewis
and S. O'Keefe for contributions to the development of the
network designs.
This work was supported by the EPSRC, the
Research Training and Development and Human Potential Programs of
the European Union, the National Security Agency (NSA) and
Advanced Research and Development Activity (ARDA)
(P-43513-PH-QCO-02107-1).

\section{Appendix: basic operations for CSS codes} \label{s:basic}

We describe the fault-tolerant implementation of the basic
gates assumed in the main text. Some of the results, such as lemmas 2
and 3 were obtained by Gottesman using stabilizer methods.
We derive them by a different method and add further information.

Consider the effect of some operation (produced by a network of
quantum gates or measurements) on the physical qubits of one or
more encoded blocks. We define an operation to be `legitimate' if
it maps the encoded Hilbert space onto itself.
{\em Transversal} application of a two-bit operator is
defined to mean the operator is applied once to each pair of
corresponding physical bits in two blocks, and similarly for
transversal three-bit operations across three blocks. Legitimate
transversal operations are fault tolerant.

Typically a legitimate transversal operation will result
in a blockwise operation (defined in section \ref{s:term}, c.f. lemma 2), but
this need not always be the case.

The tilde as in
$\tilde{U}$ is used to denote the operation $U$ applied to
the physical qubits. Operators without a tilde are understood
to act on the logical, i.e. encoded qubits.
Thus $_{L}\!\left\langle
{u}\right| {U}\left| {v} \right\rangle _{L}=\left\langle
{u}\right| \tilde{U} \left| {v}\right\rangle$.

The CSS quantum codes are those whose stabilizer generators
separate into $X$ and $Z$ parts
\cite{96:Gottesman,96:SteaneA,96:SteaneB,96:Calderbank,97:Calderbank,98:Calderbank}. We restrict
attention to these codes, rather than any stabilizer code, because
they permit a larger set of easy-to-implement fault tolerant
operations, and their coding rate $k/n$ can be close to that of
the best stabilizer codes. The CSS codes have the property that  
the zeroth quantum codeword can be written as an equal
superposition of the words of a linear classical code $C_0$,
\begin{equation}
\left| {0} \right>_L = \sum_{x \in C_0} \left| {x} \right>,
\label{cw0}
\end{equation}
where $\left| {x} \right>$ is a product state, $x$ is a binary
word ($1 \times n$ row vector), and the other codewords are formed from cosets of $C_0$. Let
$ {D}$ be the $k \times n$ binary matrix of coset leaders,
then the complete set of encoded basis states is given by
\begin{equation}
\left| {u} \right>_L = \sum_{x \in C_0} \left| {x + u  {D}} \right>,  \label{uL}
\end{equation}
where $u$ is a $k$-bit binary word ($1 \times k$ row vector).

Consider a CSS code as defined in eq. (\ref{uL}). Then one possible choice for
the encoded $X$ and $Z $ operators is
\begin{eqnarray}
{X}_u &=& \tilde{X}_{u  {D}}  \label{Xbar} \\
{Z}_u &=& \tilde{Z}_{u  {D} ( {D}^T {D})^{-1}}. \label{Zbar}
\end{eqnarray}
Equation (\ref{Xbar}) follows immediately from the code construction
(\ref{uL}). Eq. (\ref{Zbar}) may be obtained as follows.
Since we are dealing with row vectors, the scalar product is $x
\cdot y = x y^T$. Now, consider $y \in C_0^{\perp}$: then $\tilde{Z}_y
|x + u  {D} \left. \right> = (-1)^{y \cdot u  {D}} | x + u
  {D} \left. \right>$ and hence
\begin{eqnarray}
\tilde{Z}_y  \ket{u}_L &=& (-1)^{y \cdot u  {D}} \ket{u}_L \\
\mbox{ but }
{Z}_v  \ket{u}_L &=& (-1)^{v \cdot u} \ket{u}_L
\end{eqnarray}
so we need to solve $v \cdot u = y \cdot (u  {D})$ for $y$:
\begin{eqnarray}
v u ^T &=& y  {D}^T u^T   \;\;\;\; \forall u \\
\Rightarrow \;\;\;\; v &=& y  {D}^T \\
\Rightarrow \;\;\;\; y &=& v  {D} ( {D}^T  {D})^{-1}
\end{eqnarray}
where we assume the inverse of the square matrix ${D}^T {D}$ exists. We
will mostly be concerned with cases where $ {D}^T  {D}$ is an identity
matrix. To check for
consistency, we should confirm that $y \in C_0^{\perp}$ as was
assumed---the proof of this is omitted here, but it is obvious for
the case of a weakly self-dual code with $ {D}^T  {D} = I$.

Note that  , when operating on codewords, $\tilde{X}_{x+y}$ is equivalent
to $\tilde{X}_x$ for all $y \in C_0$, so each ${X}$ operator
is a member of a group of $2^{\kappa}$ equivalent operators, where
$\kappa = (n-k)/2$ is the size of $C_0$. Another way of seeing
this is to note that since $C_0 \subset C_0^{\perp}$, $\tilde{X}_{y \in
C_0}$ is in the quantum code stabilizer. Similar statements
apply to the ${Z}$ operators. The complete set of
$2^{2n}$ Pauli $\tilde{X}$ or $\tilde{Z}$ operators on $n$ bits is thus divided up
as
\begin{eqnarray*}
2^{(n-k)/2} && X\mbox{-stabilizer members} \\
2^{(n-k)/2} && Z\mbox{-stabilizer members} \\
2^k  && {X}\mbox{ operators} \\
2^k  && {Z}\mbox{ operators} \\
2^{(n-k)/2} && \mbox{detectable $X$ errors} \\
2^{(n-k)/2} && \mbox{detectable $Z$ errors}
\end{eqnarray*}

{\bf Lemma 1.} For $[[n,1,d]]$ codes where all words in $\left|
{0} \right>_L $ have weight $r_0 \mbox{ mod } w$, and all words in
$\left| {1} \right>_L$ have weight $r_1 \mbox{ mod } w$,
transversal application of the following are legitimate:
$\tilde{P}(2\pi/w)$, $^C\! \tilde{P}(4\pi/w)$, $^{CC}\!\! \tilde{P}(8\pi/w)$, and
achieve respectively ${P}(2r\pi/w)$, $^C\! {P}(4r\pi/w)$, $%
^{CC}\!\! {P}(8r\pi/w)$, where $r=r_1 - r_0$.

Lemma 1 applied to codes with $w=8$ or more provides a quicker way
to generate the Toffoli gate ${T}$ and its partners $^C\!\!
S$ and ${P}(\pi /4)$ than has been previously discovered.
The concept generalizes to $^{ccc}\!\tilde{P}(16\pi /w)$ and so on, but
the codes for which this is useful (i.e. having $w\ge 16$) are
either inefficient or too unwieldy to produce good error
thresholds.

{\em Proof:} for clarity we will take $r_{0}=0$ and $r_{1}=r$, the
proof is easily extended to general $r_{0}$. The argument for $^C\!\!
\tilde{P}(4\pi /w)$ was given in \cite{96:KnillB}, but we shall need it for
$^{CC}\!\! \tilde{P}(8\pi /w)$, so we repeat it here. Consider $^C\!\! \tilde{P}(4\pi
/w)$ applied to a tensor product of two codewords. Let $x,y$ be
binary words appearing in the expressions for the two codewords,
and let $a$ be the overlap (number of positions sharing a 1)
between $x$ and $y$. Let $|x|$ denote the weight of a word $x$.
Then $2a=|x|+|y|-|x+y|$. There are three cases to consider. First
if $x,y\in C_{0}$ then $|x|=0\mbox{ mod }w,\;|y|=0\mbox{ mod }w$
and $|x+y|=0\mbox{ mod }w$ so $2a=0\mbox{ mod }w$ from which
$a=0\mbox{ mod }w/2$. Therefore the multiplying factor introduced
by the transversal operation is $1$. If $x\in C_{0}
$ and $y\in C_{1}$ then $x+y\in C_{1}$ so $|x|=0\mbox{ mod }w,\;|y|=|x+y|=r%
\mbox{ mod }w$ so $2a=0\mbox{ mod }w$ again. If $x,y\in C_{1}$
then $x+y\in C_{0}$ so $a=r\mbox{ mod }w/2$ and the multiplying
factor is $\exp (ir4\pi /w)$. The resulting operation in the
logical Hilbert space is therefore $^C\!\! {P}(4r\pi /w)$.

Next consider $^{CC}\!\! \tilde{P}(8\pi/w)$ applied to a tensor product of
three codewords. Let $x,y,z$ be words appearing in the three
codeword expressions,
and $a,b,c$ be the overlap between $x$ and $y$, $y$ and $z$, and $z$ and $x$%
, respectively. Let $d$ be the common overlap of $x,y$ and $z$, so
\begin{equation}
|x+y+z| = |x| + |y| + |z| - 2a - 2b - 2c + 4d.
\end{equation}
There are four cases to consider. If $x,y,z \in C_0$ then $d=0
\mbox{ mod } w/4$. If $x,y \in C_0, z \in C_1$ then $|x+y+z| =
|z|$, $2a=2b=2c=0 \mbox{
mod } w$ from the argument just given, therefore $d=0 \mbox{ mod } w/4$. If $%
x \in C_0$, $y,z \in C_1$ then $x+y+z \in C_0$, $2a=2c=0 \mbox{
mod } w$ while $2b = 2r \mbox{ mod } w = |y| + |z|$ so again $d=0
\mbox{ mod } w/4$. If $x,y,z \in C_1$ then $x+y+z \in C_1$,
$2a=2b=2c=2r \mbox{ mod } w$, therefore $d=r \mbox{ mod } w/4$.
The overall effect is that of the operation $^{CC}\!\!
{P}(8r\pi/w)$. \hfill {QED}

{\bf Lemma 2.} Transversal $^C\! \tilde{X}$ is legitimate for all CSS
codes, and acts as blockwise $^C \! {X}$.

{\em Proof:} transversal $^C\! \tilde{X}$ acts as follows:
\begin{eqnarray}
^C\! \tilde{X}_{\rm tr} \left| {u} \right>_L \left| {v}
\right>_L &=& \sum_{x \in C_0} \sum_{y \in C_0} \left| {x+u   {D}} \right> \left|
{y+v  {D} + x + u  {D}} \right> \nonumber \\
&=& \sum_{x \in C_0} \sum_{y \in C_0} \left| {x+u  {D}} \right>
\left| {y+(u+v)  {D}} \right>   \nonumber \\
&=& \left| {u} \right>_L \left| {u+v} \right>_L.  \label{CXbar}
\end{eqnarray}
This is $^C\! {X}$ from each logical qubit in the first
block to the corresponding one in the second. \hfill {QED}

{\bf Lemma 3.} Transversal $\tilde{H}$ and $^C\! \tilde{Z}$ are legitimate for any
$[[n,2k_c-n,d]]$ CSS code obtained from a $[n,k_c,d]$ classical
code that contains its dual, giving the effects
  \begin{eqnarray}
\tilde{H}_{\rm tr} \left| {u} \right>_L &=& \sum_{v=0}^{2^k-1} (-1)^{u  {D}
 {D}^T v^T} \left| {v} \right>_L,                  \label{Hbar}
\end{eqnarray}
\begin{equation}
^C\! \tilde{Z}_{\rm tr} \left| {u} \right>_L \left| {v} \right>_L =
(-1)^{u  {D}  {D}^T v^T} \left| {u} \right>_L \left| {v} \right>_L. \label{CZbar}
\end{equation}

Equation (\ref{Hbar}) is a blockwise ${H}$
when $ {D}  {D}^T =  {I}$, and is a closely
related transformation when $ {D}  {D}^T \ne  {I}$.
Equation (\ref{CZbar}) is a blockwise $^C\!\! {Z}$ when
$ {D}  {D}^T =  {I}$, and a related transformation
otherwise.

{\em Proof:} transversal $\tilde{H}$ acts as follows on $\left| {u}
\right>_L$:
  \begin{eqnarray}
\tilde{H}_{\rm tr} \sum_{x \in C_0} \left| {x + u {D}} \right> &=& \sum_{y \in C_0^{\perp}} (-1)^{u  {D}
y^T} \left| {y} \right>.
  \end{eqnarray}
If $C_0^{\perp}$ contains its dual $C_0$, as required for lemma 3,
then $ {D}$ and $C_0$ together generate $C_0^{\perp}$, so
this can be written
  \begin{eqnarray}
\tilde{H}_{\rm tr} \left| {u} \right>_L &=& \sum_{v=0}^{2^k-1}
\sum_{x \in C_0} (-1)^{u  {D}  {D}^T v^T} \left| {x + v  {D}} \right> \nonumber \\
&=& \sum_{v=0}^{2^k-1} (-1)^{u  {D}
 {D}^T v^T} \left| {v} \right>_L \label{Htrans}
  \end{eqnarray}
where to simplify the power of $(-1)$ in the first equation we used the fact that $C_0$
is generated by the parity check matrix of $C_0^{\perp}$,
so $u {D}$ satisfies the parity check $x \in C_0$.

Equation (\ref{CZbar}) is proved straightforwardly by expanding
$\ket{u}_L$ and $\ket{v}_L$ as in (\ref{uL}), and then
using
$(x+uD)(y+vD)^T = u D D^T v^T \; {\rm mod} \; 2$ for all the terms
in the sum when $C_0 \subset C_0^{\perp}$.

{\bf Lemma 4.} Let $C$ be a $[n,k_c,d]$ classical code that  
contains its dual, and for which the weights of the rows of the
parity check matrix are all integer multiples of 4. Then
transversal $\tilde{S}$ is legitimate for the $[[n,2k_c-n,d]]$ CSS code
obtained from $C$, and has the effect
  \beq \tilde{S}_{\rm tr} \ket{u}_L = i^{|u    {D}|} \ket{u}_L. \eeq

The case $ {D} {D}^{T}= {I}$, which leads to a
simple effect for transversal $\tilde{H}$, also simplifies transversal
$\tilde{S}$. If $ {D} {D}^{T}= {I}$ then every row of
$ {D}$ has odd overlap with itself (i.e. odd weight) and even
overlap with all the other rows. Using an argument along similar
lines to that in the proof of lemma 1, we
deduce that the effect is the ${S}^{r}$ operator applied
to every logical qubit in the block, where $r$ is the weight of the
relevant row of $ {D}$.

{\em Proof:} We will prove lemma 4 by showing that all the quantum
codewords have $|x + u  {D}| = |u  {D}|
\mbox{ mod } 4$, so the weights modulo 4 of the components in
(\ref{uL}) depend on $u$ but not on $x$. The effect of transversal
$\tilde{S}$ will therefore be to multiply $\left| {u} \right>_L$ by the
phase factor $i^{|u  {D}|}$.

The zeroth codeword is composed from the code $C_0 = C^{\perp}$
generated by
${H}_C$, the parity check matrix of $C$. Let $y$ and $z$ be two rows
of ${H}_C$, then the conditions of the lemma guarantee $|y| = 0
\mbox{ mod } 4$ and $|z| = 0 \mbox{ mod } 4$. Furthermore, since
$C$ contains its dual,
each row of $ {H}_C$ satisfies all the checks in $ {H}_C$, so $y$
and $z$ have even overlap $2m$. Therefore $|y+z| = 4m \mbox{ mod } 4 =
0 \mbox{ mod } 4$, therefore $|x| = 0 \mbox{ mod } 4$ for all
words in $\left| {0}
\right>_L$. Next consider a coset, formed by displacing $C_0$ by the vector $%
w=u  {D}$. Since this coset is in $C$ it also satisfies
all the checks in $ {H}_C$, therefore its members have even
overlap with any $x \in C_0$. Hence if $|w| = r \mbox{ mod } 4$
then $|x + w| = r \mbox{ mod } 4$ for all the terms in the coset,
which proves the lemma.  \hfill {QED}

{\bf Lemma 5.} For CSS codes in which transversal $^C\! \tilde{Z}$ is
legitimate, transversal $^{CC}\!\! \tilde{Z}$ is legitimate when operating
on two control blocks in
the logical Hilbert space, and a target block in the space spanned by
$\ket{0 ^{\otimes n}}$, $\ket{1^{\otimes n}}$.
If transversal $^C\! \tilde{Z}$ has the effect $\left|{u} \right>_L \left| {v}
\right>_L \rightarrow
(-1)^{uv^T} \left| {u} \right>_L \left| {v} \right>_L$, then
transversal $^{CC}\!\! \tilde{Z}$ has the effect $\left| {u}
\right>_L \left| {v} \right>_L \left| {a^{\otimes n}} \right> \rightarrow
(-1)^{a(uv^T)} \left| {u} \right>_L
\left| {v} \right>_L \left| {a^{\otimes n}} \right>$, where $a = 0$ or $1$.

{\em Proof:} Consider eq. (\ref{CZbar}) and expand $\left| {u}
\right>_L \left| {v} \right>_L$ into a sum of $2n$-bit product
states $\left| {x} \right>\left| {y} \right>$. The transversal
$^C\! \tilde{Z}$ operator can only have the effect (\ref{CZbar}) if the
overlap of $x$ and $y$ is the same, modulo 2, for every term in
the sum. Therefore the transversal $^{CC}\!\! \tilde{Z}$ operator as
described in lemma 5 produces the same number of $\tilde{Z}$ operations on
the cat state, modulo 2, for every term in the corresponding
expansion, and the effect is as described. \hfill {QED}

\bibliographystyle{unsrt}
\bibliography{quinforefs}

\end{document}